\documentclass[12pt,english]{article}

\usepackage[T1]{fontenc}
\usepackage[utf8]{inputenc}
\usepackage{geometry}
\usepackage{color}
\usepackage{float}
\usepackage{multirow}
\usepackage{amsmath,bm}
\usepackage{amssymb}
\usepackage{graphicx}
\usepackage{setspace}
\usepackage[authoryear]{natbib}
\usepackage{tikz}
\usepackage{adjustbox}
\usetikzlibrary{arrows,shapes,automata,backgrounds,decorations,petri,positioning,patterns}
\makeatletter

\providecommand{\tabularnewline}{\\}

\usepackage{amsthm}\usepackage{physics}
\usepackage{gensymb}
\usepackage{flexisym}
\usepackage{tikz}
\usepackage{mathrsfs}\usepackage{rotating}\usepackage{titlesec}
\usepackage{lipsum}\usepackage{fancyhdr}
\usepackage{framed}\usepackage{chngcntr}\usepackage{blindtext}
\usepackage{epstopdf}
\usepackage{multirow}
\usepackage{colortbl}
\usepackage{pdflscape}
\usepackage{bbm}
\usepackage{xcolor}

\newcommand{\bfbeta}{\boldsymbol \beta}

\newcommand{\bfdelta}{\boldsymbol \delta}

\newcommand{\bftheta}{\boldsymbol \theta}

\newcommand{\bfphi}{\boldsymbol \phi}

\newcommand{\bfgamma}{\boldsymbol \gamma}
\newcommand{\bfsigma}{\boldsymbol \sigma}

\newcommand{\bs}{\boldsymbol s}

\newcommand{\bfTheta}{\boldsymbol \Theta}
\newcommand{\bfS}{\mathbf S}
\newcommand{\bfy}{\mathbf y}
\newcommand{\bfZ}{\mathbf Z}
\newcommand{\bfw}{\mathbf w}

\numberwithin{equation}{section}

\newcommand{\distas}[1]{\mathbin{\overset{#1}{\kern\z@\sim}}}%
\newsavebox{\mybox}\newsavebox{\mysim}
\newcommand{\distras}[1]{%
  \savebox{\mybox}{\hbox{\kern3pt\scriptstyle#1\kern3pt}}%
  \savebox{\mysim}{\hbox{\sim}}%
  \mathbin{\overset{#1}{\kern\z@\resizebox{\wd\mybox}{\ht\mysim}{\sim}}}%
}
\newcommand{\bw}{\bfw} 

\usepackage{geometry}

\usepackage{setspace}
\@ifundefined{showcaptionsetup}{}{%
 \PassOptionsToPackage{caption=false}{subfig}}
\usepackage{subfig}

\makeatother

\usepackage{authblk}
\defcitealias{NOAA}{NOAA2019}

\newcommand{\blind}{1}

\usepackage{babel}
\begin{document}

\begin{singlespace}
\if1\blind {
\title{Hierarchical Bayesian  Nearest Neighbor Co-Kriging Gaussian Process Models; an Application to Intersatellite Calibration}
\author[1]{Si Cheng}
 \author[1]{Bledar A. Konomi \thanks{Corresponding author:
Bledar A. Konomi (alex.konomi@uc.edu)} }
\author[2]{Jessica L. Matthews}
 \author[3]{Georgios Karagiannis}
\author[1]{Emily L. Kang}
\affil[1]{University of Cincinnati}
\affil[2]{North Carolina State University, Cooperative Institute for Satellite Earth System Studies (CISESS)}
\affil[3]{Durham University}
\maketitle 
\vspace{-1em}} \fi
\end{singlespace}
\begin{abstract}
\begin{singlespace}
Recent advancements in remote sensing technology and the increasing size of satellite constellations allow for massive geophysical information to be gathered daily on a global scale by numerous platforms of different fidelity.  The auto-regressive co-kriging model provides a suitable framework for the analysis of such data sets as it is able to account for cross-dependencies among different fidelity satellite outputs. However, its implementation in multifidelity large spatial data sets is practically infeasible because the computational complexity increases cubically with the total number of observations. In this paper, we propose a nearest neighbor co-kriging Gaussian process (GP) that couples the auto-regressive model and nearest neighbor GP by using augmentation ideas. Our model reduces the computational complexity to be linear with the total number of spatially observed locations. The spatial random effects of the nearest neighbor GP are augmented in a manner which allows the specification of semi-conjugate priors. This facilitates the design of an efficient MCMC sampler involving mostly direct sampling updates.  The good predictive performance of the proposed method is demonstrated in a simulation study. We use the proposed method to analyze High-resolution Infrared Radiation Sounder data gathered from two NOAA polar orbiting satellites.

\end{singlespace}
\end{abstract}
\begin{singlespace}
\end{singlespace}

\newpage{}

\section{Introduction}

Due to  the advancement of remote sensing technology and the growing size of satellite constellations, it has become increasingly common for geophysical information to be measured by numerous platforms at similar times and locations. Aging and exposure to the harsh environment of space results in sensor degradation over the
satellite's lifetime causing a decrease on performance reliability. This results in inaccuracy of the data as a true measure for long term trend analysis \citep{goldberg2011}. Generally, newer satellites with more advanced sensors  provide information of higher fidelity than older models. Different platforms often  collect large amounts of observations with varying fidelity for spatial areas that may or may not overlap or have the same spatial footprint. Over the years multiple methods have been developed with the overarching goal of enabling the intercomparison amongst satellite platforms including: use of ground-based observations and leveraging temporally stable targets (e.g. moon, desert sites, deep convective clouds) to assess satellite sensor performance and consistency \citep{chander2013, xiong2010, nrc2004}. But to date, these methods fail to account for differing fidelity levels between satellite platforms with statistical rigor.
 
 As a specific example of how the remote sensing community currently manages this challenge, we consider the  strategy applied to the high-resolution infrared radiation sounder (HIRS) which provides measurements from multiple satellite platforms. As described in \citet{cao2004, cao2005}, the intersatellite calibration of HIRS sensors is based on calculating differences between all the near-nadir overlapping points from two satellites within a period of time. These differences are assessed in separated $10$-degree brightness temperature bins as an aggregated mean, with no spatial or temporal dependencies. The current intersatellite calibration of the observations is simply a least square bias correction term based on a linear relationship of the differences in brightness temperature in the adjacent satellite (assumed to have the higher fidelity).  However, this naive approach ignores spatial dependency by assigning the same bias correction across the spatial domain, which can possibly yield misleading results. A single composite feature, which includes adequate information from
multiple data sources in space, is preferred for statistical inference.

In geostatistics, co-kriging is a suitable framework   for analysis of spatially correlated random processes
\citep{davis1983estimation,aboufirassi1984cokriging,ver1993multivariable,furrer2011aggregation,genton2015cross}.
Complex cross-covariance functions can lead to infeasible computational
complexity, even for moderate amounts of data. To address this issue, \citet{kennedy2000predicting} proposed an autoregressive co-kriging model which is simple, but yet flexible, to model complex
dependency structures. The autoregressive co-kriging framework has gained
popularity in computer experiments \citep{qian2008bayesian,han2010new,le2013bayesian,koziel2014efficient}
due to its computational convenience. Its framework fits well with
the multi-sensor geographical information system, since the hierarchy
is established based on age and technology of the sensors.  Most of
the computational benefits of autoregressive co-kriging
are lost when the multi-fidelity data are not observed in
a hierarchically nested structure. Multi-sensor geographical information
systems are usually observed irregularly in space and are hierarchically
non-nested. Recently, \citet{konomikaragiannisABTCK2019} proposed
a Bayesian augmented hierarchical co-kriging procedure which makes
the analysis of partially-nested and/or non-nested structures possible
with feasible computational cost by splitting the augmented likelihood
into conditionally independent parts.  Despite this simplification,
the method cannot be applied directly to large data
sets. Each conditional component of the likelihood requires evaluation
of the determinant as well as inversion of a large co-variance matrix.

Recently, statistical methods for large spatial data sets have received much
attention. Many of the most popular techniques
rely on low-rank approximation \citep{banerjee2008gaussian,cressie2008fixed},
approximate likelihood methods \citep{stein2004approximating,Gramacy2015},
covariance tapering methods \citep{furrer2006covariance,kaufman2008covariance,du2009fixed},
sparse structures \citep{lindgren2011explicit,nychka2015multiresolution,datta2016hierarchical},
multiple-scale approximation \citep{Sang2012,Katzfuss2016}, and lower
dimensional conditional distributions \citep{vecchia1988estimation,stein2004approximating,datta2016hierarchical,katzfuss2017general}.
A number of these methods have been generalized to handle large data
from multiple sources. For example, \citet{nguyen2012spatial,nguyen2017multivariate} have
proposed data fusion techniques based on fixed ranked kriging \citep{cressie2008fixed}.
The accuracy of this approach relies on the number of basis functions
and can only capture large scale variation of the covariance function.
When the data sets are dense, strongly correlated, and the noise effect is sufficiently small, low rank kriging techniques have difficulty accounting for small scale variation \citep{stein2014limitations}. More recently, \citet{taylor2018spatial} embedded the nearest-neighbor Gaussian process (NNGP) into a spatial factor model and used NNGP to model the resulting  independent GP processes. This method assumes that data sets from different sources follow an overlapping structure, limiting its use for real applications.

In this paper, we propose a new computationally efficient autoregressive
co-kriging method based on the nearest neighbor Gaussian process (NNGP),
which is called the nearest neighbor Co-kriging Gaussian process
(NNCGP). The proposed method is applicable to large
non-nested and irregular spatial data sets from different platforms
having varying quality. NNCGP utilizes an approximate imputation
procedure based on a nested  reference set to address large data sets of non-nested observations. This formulation allows the evaluation of the likelihood and
predictions with low computational cost, as well as allows the specification of conditional conjugate priors. Compared
to the aforementioned models, it exhibits both computational efficiency
and flexibility. This method enables the analysis of  high-resolution infrared radiation sounder (HIRS)  data sets gathered daily from two  polar orbiting satellite series (POES) of the National Oceanic and Atmospheric Administration (NOAA). We show that the proposed method is both more accurate and computationally more  efficient for these type of data sets.  Furthermore, the NNCGP model shows significant improvement in prediction accuracy over the existing NNGP approach, based on our simulation study and the analysis for the real data application.

The layout of the paper is as follows. In Section 2, we introduce a spatial representation of the  autoregressive co-kriging model. In Section 3, we introduce our proposed NNCGP as an extension of the existing autoregressive co-kriging model. 
In Section 4, we design an MCMC approach tailored to the proposed NNCGP model that facilitates parametric and predictive inference. In Section 5, we investigate the performance of the proposed procedure on a toy example. In Section 6, we apply the proposed method to HIRS data sets from two satellites, NOAA-14 and NOAA-15. Finally, we summarize our findings in Section 7.

\section{Spatial Co-kriging Gaussian Process}

We consider $T$  platforms which gather spatial observations with similar footprint.  We assume that: (a)  observations from different platforms are correlated, (b) platform $t$ provides a more accurate representation of the ground truth than platform $t-1$, and (c) prior belief about observations from a platform  can be modeled by a Gaussian process. We  refer to the observations of platform $t$  as the observations of fidelity level $t$. Let $y_{t}(s)$ denote the output at the spatial location $s$ at fidelity level $t=1,...,T$. Here, the fidelity level index $t$ runs from the least to most accurate platform. We consider that the observed output $z_{t}(s)$ at location $s$
is contaminated by additive random noise $\epsilon_{t}(\bs)\sim N(0,\tau_{t})$
with unknown variance $\tau_{t}$, and  $y_t(\bs)$ depends on the fidelity level output $y_{t-1}(\bs)$ via an autoregressive co-kriging model.
Specifically, we model the observation $z_{t}(\bs)$ as: 
\setlength{\belowdisplayskip}{5pt}\setlength{\belowdisplayshortskip}{5pt}
\setlength{\abovedisplayskip}{5pt}\setlength{\abovedisplayshortskip}{5pt}
\begin{align}
 & z_{t}(\bs)=y_{t}(\bs)+\epsilon_{t}(\bs)\nonumber \\
 & y_{t}(\bs)=\zeta_{t-1}(\bs)y_{t-1}(\bs)+\delta_{t}(\bs),\label{eq:davdsgdaf}\\
 & \delta_{t}(\bs)=\mathbf{h}_{t}^T(\bs)\boldsymbol{\beta}_{t}+w_{t}(s), \nonumber 
\end{align} 
for $t=2,\ldots,T$, and $y_{1}(\bs)=\mathbf{h}_{1}^T(\bs)\boldsymbol{\beta_{1}}+w_{1}(\bs)$. 
Here, $\zeta_{t-1}(\bs)$
and $\delta_{t}(\bs)$ represent the scale and additive discrepancies
between the output of  platforms with fidelity levels $t$ and $t-1$, and $\epsilon(\bs)$ are uncorrelated pure error terms with variance $\tau_{t}^2$. 
Moreover, $\mathbf{h}_{t}(\cdot)$  is a vector of basis
functions and $\bfbeta_{t}$
is a vector of coefficients at fidelity level $t$. 
We model, a
priori, $w_{t}(\bs)$ as Gaussian processes, mutually independent for
different $t$; i.e. $w_{t}(\cdot)\sim GP(0,C_{t}(\cdot,\cdot,\boldsymbol{\theta}_{t}))$
where $C_{t}(\cdot,\cdot,\boldsymbol{\theta}_{t})$ is a covariance
function with  parameters $\boldsymbol{\theta}_{t}$ at fidelity 
level $t$. This implies that $y_{1},\delta_{2},\dots,\delta_{T}$
are a priory mutually independent Gaussian processes.
The unknown scale discrepancy function $\zeta_{t-1}(\bs)$ is modeled
as a basis expansion  $\zeta_{t-1}(\bs|\bfgamma_{t-1})=\mathbf{g}_{t-1}(\bs)^{T}\bfgamma_{t-1}$ (usually with low degree), 
where $\mathbf{g}_{t}(\bs)$ is a vector of  basis functions and $\{\bfgamma_{t-1}\}$ is a vector of coefficients for the scale discrepancies, for $t=2,\dots,T$.   

The statistical model in \eqref{eq:davdsgdaf} is different from the co-kriging  model of \citet{kennedy2000predicting}, which was developed for the analysis of deterministic computer models, because it accounts for a nugget effect through $\epsilon(\bs)$.
The introduction of a nugget effect can play an important role by accounting for  measurement errors in spatial statistics as well as modeling the error in stochastic computer models \citep{baker2020analyzing}.
 The benefits of considering a nugget effect
in spatial data models has been previously noticed by \citet{Cressie1993}
and \citet{Stein99}. \citet{Gramacy2012StatComp} argued that the use of a nugget can also mitigate poor fitting when there is deviation from the GP model assumptions.  Finally, \citet{kennedy2000predicting} originally used a constant scalar discrepancy based on stationarity arguments. We use a more general  polynomial format for the scalar discrepancy   \citep{ZhiguangConnerJanetWu2005} for model flexibility and to improve predictions when needed. To avoid identifiability issues, it is recommended to use low degree basis expansions for both the additive and scalar discrepancy. 

For each level of fidelity, we choose a product  exponential covariance function:  $C_t(\bs,\bs'|\boldsymbol{\theta}_t)=\sigma_t^{2}\text{exp}\left(-\sum_{i=1}^{d}\frac{\left|s_{i}-s_{i}'\right|}{\phi_{t,i}}\right)$,
where $\boldsymbol{\theta}_t=\{\sigma_t^2,\boldsymbol{\phi}_t\}=\{\sigma_t^2,\phi_{t,1},\phi_{t,2},\ldots,\phi_{t,d}\}$, $\sigma_t^2$ is the variance parameter and $\phi_{t,i}$
control the spatial dependence strength in $\mathbb{R}^{d}$ at fidelity level $t$ and direction $i$. The product covariance function is equivalent to a diagonal anisotropic covariance function. More intricate covariance functions, such as the stationary ones from the Mat{\'e}rn family  \citep{Cressie1993, Stein99, BanerjeeGelfand2014} or the non-stationary ones of \citep{Paciorek06,Konomi2014JCGS} can also be used in this model. 

Let's assume the system is observed at $n_t$ locations at each fidelity level $t$. Let $\bfS_{t}=\{\bs_{t,1},\dots,\bs_{t,n_t}\}$ be the set of $n_t$ observed locations and  $\bfZ_{t}={z}_{t}(\bfS_t)=\{z_{t}(\bs_{t,1}),\dots,z_{t}(\bs_{t,n_{t}})\}$ represent
the observed outputs at fidelity level $t$. The joint sampling  distribution of the observations at all levels
$\bfZ_{1:T}=\{\bfZ_{1},\ldots,\bfZ_{T}\}$ is Gaussian, hence the likelihood  $L(\bfZ_{1:T}|\bftheta_{1:T},\bfbeta_{1:T},\bfgamma_{1:{T-1}},\tau_{1:T})$ is a multivariate Normal density function with mean vector $\boldsymbol{\mu}$ and covariance matrix $\boldsymbol{\Lambda}$ that cannot  easily be computed. Specifically, if the data are observed in non-nested  locations for each fidelity level, the calculation of the likelihood requires $\mathcal{O}((\sum_{t=1}^{T}n_t)^3)$ flops to invert the covariance matrix $\boldsymbol{\Lambda}$ and an additional $\mathcal{O}((\sum_{t=1}^{T}n_t)^2)$ memory to store it as explained in \cite{konomikaragiannisABTCK2019}. Thus the likelihood evaluation  is computationally costly, if not practically impossible, when $n_t$ is large. For instance, in our application, for each satellite we have $\sim 10^5$  observations at  non-nested locations making the practical implementation impossible.  

\section{ Nearest Neighbor Co-kriging Gaussian Process}

To deal with the computational complexity of the co-kriging model,  we propose to apply a set of independent nearest-neighbor Gaussian process (NNGP) priors \citep{datta2016hierarchical} at the spatial process of each level of fidelity. NNGP is a fully dimensional GP with sparse representation in the precision matrix of the spatial process.  Let  $\bfw_{t}=w_t(\bfS_t)=\{w_{t}(\bs_{t,1}),\ldots,w_{t}(\bs_{t,n_{t}})\}$ denote the vector of the spatial process over the observed locations $\bfS_{t}$ at fidelity level $t$.  Based on the independent assumptions in \eqref{eq:davdsgdaf}, the joint density of $\bw_{1:T}$ can be written as the product of conditional Normal densities:
\begin{align}
p(\bw_{1:T}|\bftheta_{1:T})=\prod_{t=1}^{T}p(\bfw_{t}|\bftheta_{t})=\prod_{t=1}^{T}\prod_{i=1}^{n_{t}}p(w_{t}(\bs_{t,i})|\bfw_{t,<i}),
\end{align}
where $p(\bfw_{t}|\bftheta_{t})=N(\bfw_{t}|0,\mathbf{C}(\boldsymbol{\theta}_{t}))$ , $\prod_{i=1}^{n_{t}}p(w_{t}(\bs_{t,i})|\bfw_{t,<i})$ is the conditional representation of the joint distribution of $\bfw_{t}$, $\bfw_{t,<i}=\{w_{t}(s_{t,1}),w_{t}(\bs_{t,2}),\ldots,w_{t}(\bs_{t,i-1})\}$
for $2\leq i\leq n_{t}$, and $\bfw_{t,<1}=\emptyset$.   We specify a multivariate Gaussian distribution over a fixed set of points in the domain, to which we refer to as the \textit{reference set}. For simplicity and computational efficiency, the reference set is chosen to coincide with the set of observed locations $\bfS_t$.  As demonstrated in \citep{datta2016hierarchical},  based on the reference set, we can extend this finite-dimensional multivariate Normal distribution to a stochastic process over the domain.  

Let $N_{t}(\bs_{t,i})$ be a subset of locations from $\bfS_{t,<i}=\{\bs_{t,1},\bs_{t,2},\ldots,\bs_{t,i-1}\}$. $N_{t}(\bs_{t,i})$ is constructed by choosing at most $m$ ``nearest neighbors'' of location $\bs_{t,i}$ from $\bfS_{t,<i}$ such that: 
\begin{align}
N_{t}(\bs_{t,i})=\begin{cases}
\emptyset & ,\text{ for } i=1,\nonumber\\
\{\bs_{t,1},\bs_{t,2},\ldots,\bs_{t,i-1}\} & ,\text{ for }\ 2\leq i\leq m,\nonumber\\
\text{$m$ nearest neighbors among}\{\bs_{t,1},\bs_{t,2},\ldots,\bs_{t,i-1}\} & ,\text{ for } i>m.
\end{cases}
\end{align}
 Given the above specification of nearest neighbors, and its ordering mechanism, the density  $p(\bfw_{t}|\bftheta_{t})$ is approximated by   $\tilde{p}(\bfw_{t}|\bftheta_{t})=\prod_{i=1}^{n_{t}}p(w_{t}(\bs_{t,i})|\bfw_{t,N_{t}(\bs_{t,i})})$. It can be shown that $w_{t}(\bs_{t,i})|\bfw_{t,N_{t}(\bs_{t,i})}\sim N(\mathbf{B}_{t,\bs_{t,i}}\bfw_{t,N_{t}(\bs_{t,i})},F_{t,\bs_{t,i}})$,
where $\mathbf{B}_{t,\bs_{t,i}}=\mathbf{C}_{\bs_{t,i},N_{t}(\bs_{t,i})}^T\mathbf{C}_{N_{t}(\bs_{t,i})}^{-1}$,
$F_{t,\bs_{t,i}}=\mathbf{C}(\bs_{t,i},\bs_{t,i})-\mathbf{C}_{\bs_{t,i},N_{t}(\bs_{t,i})}^T\mathbf{C}_{N_{t}(\bs_{t,i})}^{-1}\mathbf{C}_{\bs_{t,i},N_{t}(\bs_{t,i})}$,
$\mathbf{C}_{\bs_{t,i},N_{t}(\bs_{t,i})}$ is the covariance matrix of
$w_{t}(\bs_{t,i})$ and $\bfw_{t,N_{t}(\bs_{t,i})}$, and $\mathbf{C}_{N_{t}(\bs_{t,i})}$
is the covariance matrix of $\bfw_{t,N_{t}(\bs_{t,i})}$. Thus the nearest neighbor density $\tilde{p}(\bfw_{t}|\boldsymbol{\theta}_t)$ is Normal with mean zero and covariance $\mathbf{\tilde{C}}(\boldsymbol{\theta}_t)$,
where $\mathbf{\tilde{C}}^{-1}(\boldsymbol{\theta}_t)$
is a sparse matrix with at most $\frac{1}{2}n_{t}m(m+1)$ non-zero
elements (Appendix \ref{App1}). 

With  NNGP prior specification for $\bfw_t$ and general prior formulation for $\bfTheta_{t}=(\bfphi_{t}, \bfsigma^2_{t},\bfbeta_{t},\bfgamma_{t-1},\tau_{t})$, the posterior distribution is 
\begin{align}
    p(\bfTheta_{1:T},\bfw_{1:T}| & \bfZ_{1:T})
    \propto  
    L(\bfZ_{1:T}|\bftheta_{1:T},\bfbeta_{1:T},\bfgamma_{1:{T-1}},\tau_{1:T},\bfw_{1:T})
    \  
    \prod_{t=1}^{T} 
    p(\bfTheta_{t})\tilde{p}(\bfw_{t}|\bftheta_{t})
    \label{eq:dsvsgsdbsbg}
\end{align}
while the likelihood kernel is a multivariate Normal density with mean $\boldsymbol{\mu}$ and covariance $\boldsymbol{\Lambda}$, which are defined in Appendix B. The covariance matrix $\boldsymbol{\Lambda}$ is not sparse since the cross covariance in  $\bfS_{t'}\backslash\bfS_{t}$  for $t'>t$ is not zero. So the likelihood, conditional on the nearest neighbor spatial random effects,  cannot be simplified  unless $\bfS_{t'}\backslash\bfS_{t}=\emptyset$ (or equivalently $\bfS_{t'}\subseteq\bfS_{t}$, that is the observation locations are in a nested hierarchical structure).   Thus, the direct implementation of NNGP on  $\bw_{1:T}$  when observed locations are not fully nested (such that $\bfS_{t}\backslash\bfS_{t'}=\emptyset$ for $t>t'$) may still lead to infeasible computational complexity. 

 To overcome the computational issue, we introduce new evaluations of the spatial  process $w_{t}(\cdot)$ for each level. The new evaluations of  $w_{t}(\cdot)$  are done in such a way  that the reference set of level $t$ is nested within the reference set of level $t+1$. Choosing this fully nested structure allows the likelihood and the posterior to be factorized into $T$ conditionally independent parts. Moreover, each of the conditionally independent parts of the factorized likelihood has a diagonal covariance matrix similar to NNGP. We call this new computationally efficient procedure the Nearest Neighbor Co-kriging Gaussian Process (NNCGP). NNCGP utilizes the computational advantages of both the auto-regressive co-kriging model and the NNGP.  The proposed NNCGP is a well-defined process derived from a parent co-kriging Gaussian process. For any fidelity level $t$ and finite set $V\in D$,   $\tilde{p}(w_{t,V})$ is the density of the realizations of a Gaussian process over $V$.

Consider observed data sets $\{\bfZ_{t},\bfS_{t}\}$, with the corresponding
spatial process vectors and output vectors  $\mathbf{y}_{t}=y_{t}(\bfS_t)=\{y_{t}(\bs_{t,1}),\ldots,y_{t}(\bs_{t,n_{t}})\}$.
Set $\bfS_{t}^{*}=\bigcup\limits _{i=t+1}^{T}\bfS_{i}\backslash\bfS_{t}=\{s_{t,1}^{*},\ldots,s_{t,n_{t}^{*}}^{*}\}$
as an additional reference set of fidelity level $t$, which contains the observed
locations that are not in the $t^{th}$ level but are in higher fidelity levels. Denote $\bfw_{t}^{*}=\{w_{t}(\bs_{t,1}^{*}),\ldots,w_{t}(\bs_{t,n_{t}^{*}}^{*})\}$
as the spatial interpolants with corresponding $\mathbf{y}_{t}^{*}$. By construction, $S_t^*$  is the smallest collection of sets of spatial
locations required to be added to the original observations $S_t$ in order to obtain hierarchically
nested locations. Consequently, $S_1^*$ consists of all locations observed in higher fidelity levels but not at the first fidelity level and  $S_T^*=\emptyset$.
Let $\tilde{\bfw}_{t}=\bfw_{t}^{*}\cup\bfw_{t}$, $\mathbf{\tilde{y}}_{t}=y_{t}(\bfS_{t}^{*})\cup y_{t}(\bfS_{t})$,
$\boldsymbol{\tilde{S}}_{t}=\bfS_{t}^{*}\cup\bfS_{t}$, and $\tilde{n}_{t}=n_{t}+n_{t}^{*}$.
Thus the complete set of observed locations $\boldsymbol{\tilde{S}}_{t}$
and $\bfS_{t'}$ have a nested hierarchical structure with $\bfS_{t'}\subseteq\boldsymbol{\tilde{S}}_{t}$
when $t'\geq t$. By sequentially adding $\bfw_{t}^{*}$ to each level, we can construct a fully nested hierarchical model. Figure 1 illustrates the proposed procedure in a  directed  acyclic  graph (DAG) representation of a toy example with two fidelity levels. 

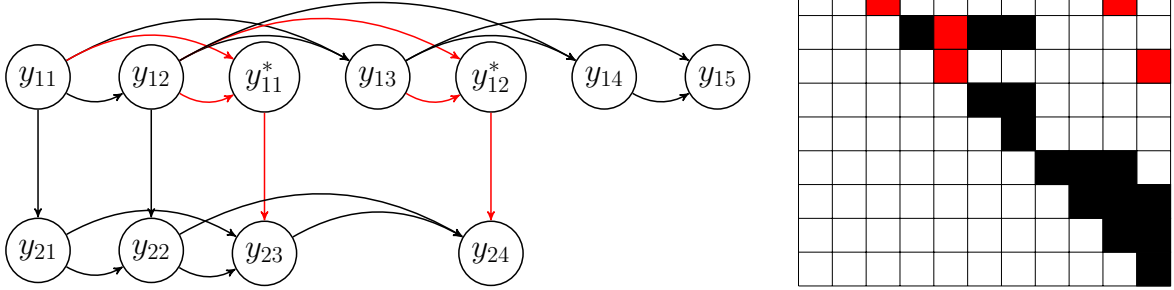
\begin{figure}
\centering
\begin{adjustbox}{width=0.6\textwidth}

\begin{tikzpicture}[->,>=stealth',auto,node distance=2.1cm,
  thick,main node/.style={circle,draw,font=\sffamily\Large\bfseries}]

  \node[main node] (1) {$y_{11}$};
  \node[main node] (2) [right of=1] {$y_{12}$};
  \node[main node] (3) [right of=2] {$y^{*}_{11}$};
  \node[main node] (4) [right of=3] {$y_{13}$};
  \node[main node] (5) [right of=4] {$y^{*}_{12}$};
  \node[main node] (6) [right of=5] {$y_{14}$};
  \node[main node] (7) [right of=6] {$y_{15}$};

  \node[main node] (15) [below = 2cm of 1] {$y_{21}$};
  \node[main node] (16) [below = 2cm of 2] {$y_{22}$};
  \node[main node] (17) [below = 2cm of 3] {$y_{23}$};
  \node[main node] (19) [below = 2cm of 5] {$y_{24}$};

 \path[every node/.style={font=\sffamily\small}]
  
  (1) edge node [below] {} (15)
  (2) edge node [below] {} (16)
  (3) edge [red] node [below] {} (17)
  (5) edge [red] node [below] {} (19)

    (1) edge [bend right] node [right] {} (2)
    (2) edge [bend right,red] node [right] {} (3)
    (4) edge [bend right,red] node [right] {} (5)
    (6) edge [bend right] node [right] {} (7)
    (1) edge[bend left] node [right] {} (4)
    (1) edge[bend left, red] node [right] {} (3)
    (2) edge[bend left] node [right] {} (4)
    (2) edge[bend left,red] node [right] {} (5)
    (2) edge[bend left] node [right] {} (6)
    (4) edge[bend left] node [right] {} (6)
    (4) edge[bend left] node [right] {} (7)

    (15) edge [bend right]  node [right] {} (16)
    (16) edge [bend right]  node [right] {} (17)

    (15) edge[bend left] node [right] {} (17)
    (16) edge[bend left] node [right] {} (19)
    (17) [bend left] edge node [right] {} (19)
    ;  
    ;
    \node[above = 0.3cm,font=\Large\bfseries] at (current bounding box.north) {Graphical representation of NNCGP};
\end{tikzpicture}
\end{adjustbox}\hspace{0.5cm}
\begin{adjustbox}{width=0.3\textwidth}
\begin{tikzpicture}
\draw[step=1cm,black,thin] (0,0) grid (11,-11);

\filldraw[fill=black, draw=black] (0,0) rectangle (2,-1);
\filldraw[fill=red, draw=black] (2,0) rectangle (3,-1);
\filldraw[fill=black, draw=black] (3,0) rectangle (4,-1);
\filldraw[fill=black, draw=black] (7,0) rectangle (8,-1);

\filldraw[fill=black, draw=black] (1,-1) rectangle (2,-2);
\filldraw[fill=red, draw=black] (2,-1) rectangle (3,-2);
\filldraw[fill=black, draw=black] (3,-1) rectangle (4,-2);
\filldraw[fill=red, draw=black] (4,-1) rectangle (5,-2);
\filldraw[fill=black, draw=black] (5,-1) rectangle (6,-2);
\filldraw[fill=black, draw=black] (8,-1) rectangle (9,-2);

\filldraw[fill=red, draw=black] (2,-2) rectangle (3,-3);
\filldraw[fill=red, draw=black] (9,-2) rectangle (10,-3);
\filldraw[fill=black, draw=black] (3,-3) rectangle (4,-4);
\filldraw[fill=red, draw=black] (4,-3) rectangle (5,-4);
\filldraw[fill=black, draw=black] (5,-3) rectangle (7,-4);
\filldraw[fill=red, draw=black] (4,-4) rectangle (5,-5);
\filldraw[fill=red, draw=black] (10,-4) rectangle (11,-5);
\filldraw[fill=black, draw=black] (5,-5) rectangle (7,-6);
\filldraw[fill=black, draw=black] (6,-6) rectangle (7,-7);
\filldraw[fill=black, draw=black] (7,-7) rectangle (10,-8);
\filldraw[fill=black, draw=black] (8,-8) rectangle (10,-9);
\filldraw[fill=black, draw=black] (10,-8) rectangle (11,-9);

\filldraw[fill=black, draw=black] (9,-9) rectangle (10,-10);
\filldraw[fill=black, draw=black] (10,-9) rectangle (11,-10);
\filldraw[fill=black, draw=black] (10,-10) rectangle (11,-11);

\end{tikzpicture}
\end{adjustbox}

\caption{Toy  example  of the NNCGP for two fidelity levels (T=2) with $n_1=5$, $n_2=4$, $n_1^*=2$.  Left:    directed  acyclic  graphical representation of  the noiseless  part of the NNCGP model. Right:  covariance matrix sparsity of the model (white squares represent zero cross-covariance).}
\end{figure}

Using the Markovian property of the co-kriging model, the joint
likelihood can be factorized as a product of likelihoods from different fidelity levels conditional on augmented spatial interpolants,  i.e.: 
\begin{align}
L(\boldsymbol{Z}_{1:T}|\cdot) & =p(\bfZ_{1}|\bfw_{1},\bfbeta_{1},\tau_{1})\prod_{t=2}^{T}p(\bfZ_{t}|\bfw_{t},\bfbeta_{t},y_{t-1}(\bfS_{t}),\bfgamma_{t-1},\tau_{t})\nonumber \\
 & =N(\bfZ_{1}|\mathbf{h}_{1}(\bfS_{1})\bfbeta_{1}+\bfw_{1},\tau_{1}\mathbf{I})\prod_{t=2}^{T}N(\bfZ_{t}|\zeta_{t-1}(\bfS_{t})\circ y_{t-1}(\bfS_{t})+\delta_{t}(\bfS_{t}),\tau_{t}\mathbf{I}),\label{conditional_ind}
\end{align}
where $\circ$ is the Hadamard production symbol and  $y_{t-1}(\bfS_{t}) \subset \mathbf{\tilde{y}}_{t-1}$.  Based on the NNGP prior on the $\bfw_{1:T}$ described above, we can write the joint prior distribution of $\tilde{\bfw}_{1:T}$ as:
\begin{align}
\tilde{p}(\tilde{\bfw}_{1:T}|\bftheta_{1:T}) & = \prod_{t=1}^{T} \tilde{p}(\bfw^*_{t}|\bfw_{t},\bftheta_{t})\tilde{p}(\bfw_{t}|\bftheta_{t}).\label{conditional_indW}
\end{align}
Given the above representation of the likelihood and prior,
the joint posterior density function of NNCGP for a $T$ level system is:
\begin{align}
p(\boldsymbol{\Theta}_{1:T},&\tilde{\bfw}_{1:T}|\bfZ_{1:T}) \propto  p(\boldsymbol{\Theta}_{1})\tilde{p}(\bfw_{1}|\boldsymbol{\theta}_{1})\tilde{p}(\bfw_{1}^{*}|\bfw_{1},\boldsymbol{\theta}_{1})N(\bfZ_{1}|\mathbf{h}_{1}(\bfS_{1})\bfbeta_{1}+\bfw_{1},\tau_{1}\mathbf{I})\nonumber \\
 & \times\prod_{t=2}^{T}\left\{ p(\boldsymbol{\Theta}_{t})\tilde{p}(\bfw_{t}|\boldsymbol{\theta}_{t})\tilde{p}(\bfw_{t}^{*}|\bfw_{t},\boldsymbol{\theta}_{t})N(\bfZ_{t}|\zeta_{t-1}(\bfS_{t})\circ y_{t-1}(\bfS_{t})+\delta_{t}(\bfS_{t}),\tau_{t}\mathbf{I})\right\} .\label{joint_NNCGP_dist}
\end{align}
The computational complexity of implementing the NNCGP model is
dominated by the evaluation and storage of $T$ sparse matrices $(\mathbf{\tilde{C}}_{1}^{-1}(\bftheta_{1}),\dots,\mathbf{\tilde{C}}_{T}^{-1}(\bftheta_{T}))$.
Thus, the joint posterior distribution of the NNCGP model can be calculated
using $\mathcal{O}(\sum_{t=1}^T\tilde{n}_{t}m^{3})$ flops and needs
$\mathcal{O}(\sum_{t=1}^T\tilde{n}_{t}m^{2})$ dynamic memory storage. Introducing $\bfw_{1:T}^{*}$
reduces the computational complexity
as well as enables the specification of semi-conjugate priors for $ (\bfsigma^2_{t},\bfbeta_{t},\bfgamma_{t-1},\tau_{t})$ which facilitates tractability of posterior marginals and conditionals, as we explain below. An alternative approach is to  use a common reference set for all of the fidelity levels. However, this may result into a more expensive procedure since each level will have the same computational complexity as the first level $\mathcal{O}(T\tilde{n}_{1}m^{3})$.

\section{Bayesian Inference}\label{sec:GibbsSampler}

In this section, we present the MCMC sampler for the inference of parameters $\boldsymbol{\Theta}_{1:T}$
for a $T$ level NNCGP with observations $\bfZ_{1:T}$
and spatial location input sets $\bfS_{1:T}$. We also present the
prediction procedure for output $\bfZ_{t}(\bs_{p})$ at an unobserved location $\bs_{p}$
for any specified fidelity level $t$.

NNCGP model representation allows us to construct an efficient MCMC sampler to facilitate parameter and prediction inference. Since the components of $\bfw_{t}^{*}|\bfw_{t}$
are independent, we can update $\bfw_{t}^{*}$ individually. For locations $\bs_{u}\in\bfS_{t}^{*}$, the full conditional  posterior distribution of $w_{t}(\bs_{u})$ $\sim$ $ N(V_{t}^*(\bs_u)\mu_{t}^*(\bs_u),V_{t}^*(\bs_u))$
with  $V_{t}^*(\bs_u)$ and $\mu_{t}^*(\bs_u)$  are specified in Equation C3 in the Appendix. The introduction of the spatial interpolant $\bfw_{t-1}^{*}$ provides the full conditional posterior distribution of a spatial random process $\bfw_{t}$ as 
$$w_{t}(\bs_{u})|\bfbeta_{t}  ,\bftheta_{t},\tau_{t},\bfZ_t,\tilde{\bfy}_{t-1},\bfgamma_{t-1} \sim N(V_{\bfw_{t}}(\bs_{u})\mu_{\bfw_{t}}(\bs_{u}),V_{\bfw_{t}}(\bs_{u}))
$$ 
where 
\begin{align}
V_{\bfw_{t}}(\bs_{u}) & =  \left(F_{t,\bs_{u}}^{-1}+\tau_{t}^{-2}\right)^{-1},\nonumber \\
\mu_{\bfw_{t}}(\bs_{u}) &=  \tau_{t}^{-2}[z_{t}(\bs_{u})-\mathbf{h}_{t}^T(\bs_{u})\boldsymbol{\beta}_{t}-\zeta_{t-1}(\bs_{u})y_{t-1}(\bs_{u})]+F_{t,\bs_{u}}^{-1}\mathbf{B}_{t,\bs_{u}}\bfw_{t,N_{t}(\bs_{u})},\label{w_sampler}
\end{align}
for $t=2,\dots,T$, $\bs_{u}\in \bfS_{t}$. The full conditional distribution of $\bfw_1$ is similar  to the univariate case of the full conditional distribution of the spatial process \citep{datta2016hierarchical},  see Equation \ref{w_sampler} in the Appendix.

To take full advantage of the posterior representation in \eqref{joint_NNCGP_dist},  we chose independent  prior distributions for  parameters at different levels such as:
\begin{align}
p(\boldsymbol{\Theta}_{1:T})=\prod_{t=1}^{T} p(\sigma_{t})p(\boldsymbol{\phi}_{t})p(\boldsymbol{\beta}_{t})p(\tau_{t})p(\bfgamma_{t-1}).
\end{align}
The above prior representation coupled with \eqref{joint_NNCGP_dist} results in $T$  separate conditional parts for the posterior. To facilitate further computations, we assign conditional conjugate priors:     
$\sigma_{t}^{2}\sim IG(a_{t},b_{t}),\ \boldsymbol{\beta}_{t}\sim N(\boldsymbol{\mu}_{\beta_{t}},\mathbf{V}_{\boldsymbol{\beta}_{t}}),\ \tau_{t}^{2}\sim IG(c_{t},d_{t})$ for $ t=1,2,\ldots,T$ and $ \bfgamma_{t-1}\sim N(\mu_{\bfgamma_{t-1}},V_{\bfgamma_{t-1}})$ for $\ t=2,3,\ldots,T$, which leads to standard full conditional posteriors \allowdisplaybreaks
\begin{align}
 \boldsymbol{\beta}_{t}|\bfw_{t},\tilde{\bfy}_{t-1},\bfgamma_{t-1},\tau_{t},\bfZ_{t} & \sim N(\mathbf{V}_{\boldsymbol{\beta}_{t}}^{*}\boldsymbol{\mu}_{\boldsymbol{\beta}_{t}}^{*},\mathbf{V}_{\boldsymbol{\beta}_{t}}^{*}),\nonumber \\
 \bfgamma_{t}|\tilde{\bfy}_{t},\boldsymbol{\beta}_{t+1},\tau_{t+1},\bfZ_{t+1} & \sim N(\mathbf{V}_{\bfgamma_{t}}^{*}\boldsymbol{\mu}_{\bfgamma_{t}}^{*},\mathbf{V}_{\bfgamma_{t}}^{*}),\nonumber \\
 \sigma_{t}^{2}|\tilde{\bfw}_{t},\boldsymbol{\phi}_{t} & \sim IG(a_{\sigma_{t}}^{*},b_{\sigma_{t}}^{*}),\nonumber \\
\tau_{t}^2|\boldsymbol{\beta}_{t},\bfw_{t},\tilde{\bfy}_{t-1},\bfZ_{t},\bfgamma_{t-1}  & \sim IG(a_{\tau_{t}},b_{\tau_{t}}),\label{gibbs_sampler}
\end{align}
where the parameters are specified in Equations  \ref{gibbs_sampler}-\ref{eq.gamma_t} of the Appendix. For the range parameter $p({\phi}_{t,j})$, we chose the bounded prior $p({\phi}_{t,j}),\sim U(0,l_{t,j})$ to avoid numerical instabilities,
where $l_{t,j}$ is defined from the researcher and is usually associated
with the maximum distance in the $j^{th}$ direction. The conditional posterior distribution for $\boldsymbol{\phi_{t}}$
in NNCGP model is {\small{}{}{} 
\begin{align}
p(\boldsymbol{\phi}_{t}|\tilde{\bfw}_{t},\sigma_{t}^2)\propto p(\boldsymbol{\phi}_{t})|\mathbf{\tilde{C}}(\boldsymbol{\theta}_{t})|^{-1/2}\text{exp}\left\{ -\frac{1}{2}\tilde{\bfw}_{t}^{T}\mathbf{\tilde{C}}^{-1}(\boldsymbol{\theta}_{t})\tilde{\bfw}_{t}\right\}, \label{phi_sampling}
\end{align}
}where $\boldsymbol{\phi}_{t}$ appears in the sparse  cross covariance matrix
$\mathbf{\tilde{C}}(\bftheta_{t})$ and it cannot be sampled directly. The Metropolis-Hasting algorithm \citep{hastings1970monte}
can be used to update $\boldsymbol{\phi}_{t}$ in the full conditional
distribution.

For a new input location $s_{p}\not\in\boldsymbol{\tilde{S}}_{t}$, the
prediction process is to generate $z_{t}(s_{p})$ based on its predictive
distribution. Subsequently, we generate $w_{t'}(s_{p})$ independently for each level from sampler $w_{t'}(\bs_{p})\sim N(V_{t',\bs_{p}}\mu_{t',\bs_{p}},V_{t',\bs_{p}})$ for $t' = 1,...,t$; where $V_{t',\bs_{p}}$, $\mu_{t',\bs_{p}}$ are specified in \eqref{eq.w_prediction}, while $y_{t'}(\bs_p)$ are generated by $y_{t'}(\bs_p) = \zeta_{t'-1}(\bs_{p})y_{t'-1}(\bs_{p})+\mathbf{h}_{t'}(\bs_{p})\boldsymbol{\beta}_{t'}+w_{t'}(\bs_p)$. The $z_{t}(\bs_{p})$ is generated by the MCMC sampler $z_{t}(\bs_{p})|\cdot\cdot\cdot  \sim N(y_{t}(\bs_{p}),\tau_t)$.

\section{Synthetic data example\label{sec:Synthetic-data-example}}

 This section conducts a simulation study to evaluate the performance of the proposed NNCGP model in comparison to the NNGP model using the highest fidelity level data only (denoted as the single level NNGP model) and using both fidelity level data sets combined as a single data-set (denoted as the combined NNGP model).  Details about the metrics used for the comparison can be found in Appendix E. The simulations were performed in MATLAB R2018a, on a computer with specifications (intelR i7-3770 3.4GHz Processor, RAM 8.00GB, MS Windows 64bit).

\begin{figure}
\centering\subfloat[Low-fidelity observations\label{fig:sfig1-7}]{\centering \includegraphics[width=0.33\linewidth]{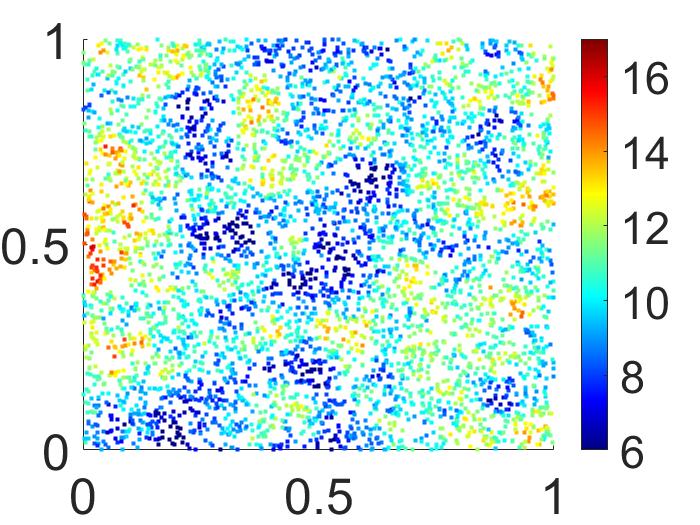}
}\subfloat[High-fidelity observations\label{fig:sfig2-5}]{\centering \includegraphics[width=0.33\linewidth]{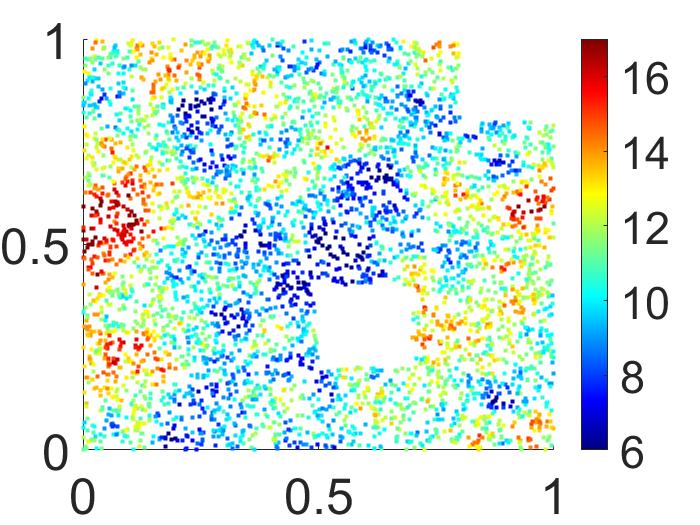}
}\subfloat[High-level testing data\label{fig:sfig1}]{\centering\includegraphics[width=0.33\linewidth]{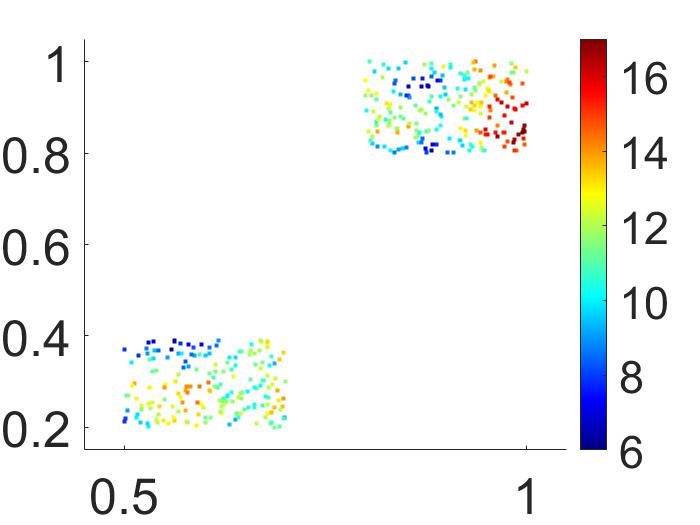}}\hfill
\subfloat[\label{fig:sfig1-8}NNCGP prediction]{\centering \includegraphics[width=0.33\linewidth]{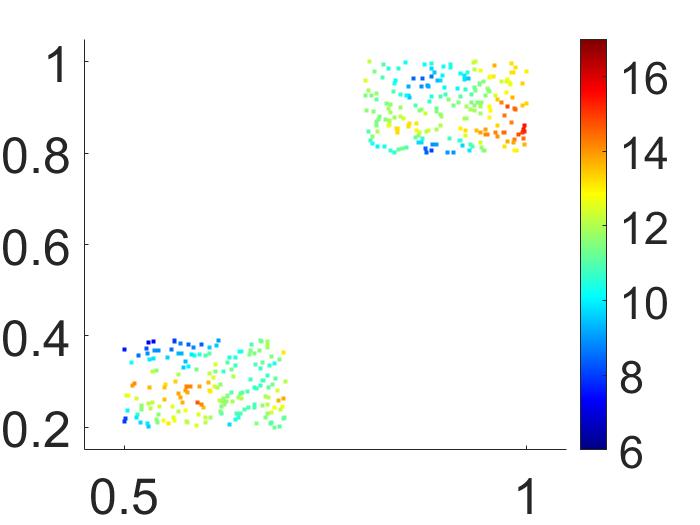}
}\subfloat[Single level NNGP prediction\label{fig:sfig2}]{\centering \includegraphics[width=0.33\linewidth]{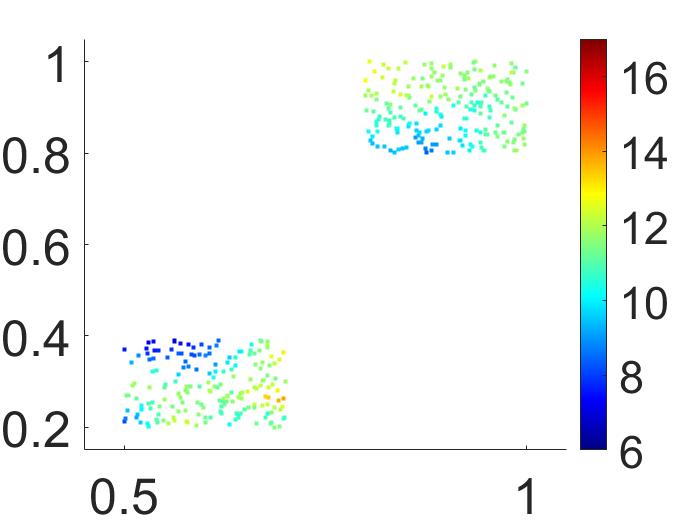}
}\subfloat[Combined NNGP prediction\label{fig:sfig2}]{\centering \includegraphics[width=0.33\linewidth]{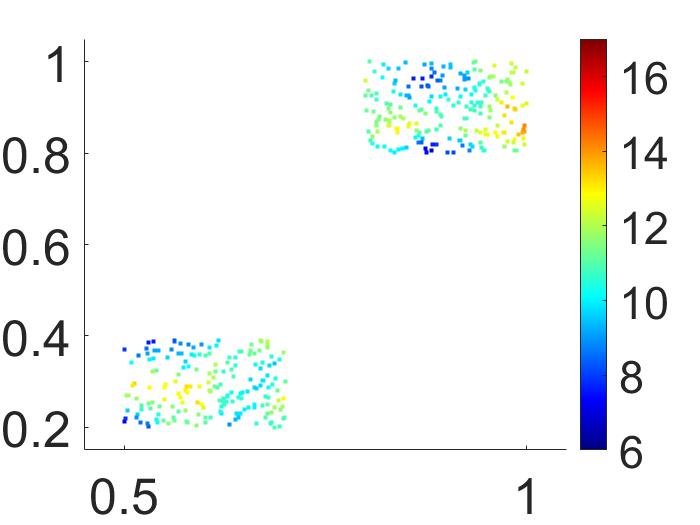}}\caption{Non-nested observations with a two (low- and high-) fidelity
level structure. White boxes in (b) indicate the testing regions. Original testing data (c) along with predictions of the high fidelity
level data-set by the (d) multifidelty Nearest Neighbor Co-Krigging
Gaussian Process (NNCGP), (e) single level nearest neighbor
Gaussian process (NNGP), and (f) the combined nearest neighbor
Gaussian process (NNGP) under the  non-nested structure.{\large{}{}\label{fig:fig-4}}}
\end{figure}

We consider a two-fidelity level system in a two dimensional unit square domain, parameterized as an auto-regressive co-kriging Gaussian process as specified in \eqref{eq:davdsgdaf}. For simplicity, the mean of $y_{1}(s)$,  the mean of the additive discrepancy $\delta_{2}(s)$, and the scalar discrepancy  $\zeta_{1}(s)$ are assumed to be constant. The covariance function of  $y_{1}(s)$ and $\delta_{2}(s)$ are assumed to be exponential.  The true values of the model parameters are listed in the first column of Table \ref{tab:Univariate-two-fildelity}. Based on the above statistical model, we generated $5,000$ observations for each  fidelity level $\bfZ_{1}$ and $\bfZ_{2}$ at randomly selected locations $\bfS_{1}$ and $\bfS_{2}$ such that $\bfS_{1}\cap\bfS_{2}=\emptyset$. The $10,000$ generated observations are shown in Figures \ref{fig:fig-4}(a-c).  To assess predictive performance, we randomly selected two small square subregions for testing from the high fidelity level dataset  (Figure \ref{fig:sfig2-5}).

For the Bayesian inference of NNCGP on the unknown parameters   $\beta_{1}, \beta_{2}$ and
$\gamma_{1}$,  we assigned independent Normal prior distributions with zero mean and large variances.  We used inverse Gamma priors for the spatial and noise variances $\sigma_{t}^{2}\sim IG(2,1)$ and $\tau_{t}^{2}\sim IG(2,1)$, respectively. The range correlation parameters  $\phi_{1}$ and $\phi_{2}$ each used a uniform prior $U(0,100)$. Similar non-informative priors were used for both the single level NNGP as well as the combined NNGP model. For all three models, we ran the Markov chain Monte Carlo (MCMC) sampler as described in Section $4$ with $40,000$ iterations where the first $5,000$
iterations were discarded as burn-in. The convergence of the MCMC sampler for each parameter was assessed from their associated trace plots.

In Table \ref{tab:Univariate-two-fildelity}, we report the Monte Carlo estimates of the posterior means and the associated $95\%$ marginal credible intervals of the unknown parameters using the three different NNGP based procedures with $m=10$ neighbors. All but $\tau_1^2$ true values of the parameters are successfully included in the $95\%$ marginal credible intervals. The introduction of spatial interpolants may have caused a slight over estimation of $\tau_{2}$, however,  the true values of the nugget variances are successfully captured in the  $95\%$ marginal credible intervals. The uncertainty in the parameter estimations can be improved with a semi-nested or nested structure between the observed locations for the different fidelity levels, as shown for the auto-regressive co-kriging model in \citet{konomikaragiannisABTCK2019}.

\begin{table}
\centering
{\footnotesize{}{}\centering}%
\begin{tabular}{c|c|cl|cl|cl}
\hline 
\multirow{1}{*}{} & \multirow{1}{*}{{\footnotesize{}{} }}  & \multicolumn{6}{c}{{\footnotesize{}{} Model}}\tabularnewline
  & {\footnotesize{}{}True} & \multicolumn{2}{c}{{\footnotesize{}{}Single level NNGP}} & \multicolumn{2}{c}{{\footnotesize{}{}Combined NNGP}} & \multicolumn{2}{c}{{\footnotesize{}{}NNCGP}}\tabularnewline
\hline 
{\footnotesize{}{}$\beta_{1}$}  & {\footnotesize{}{}10}  &  {\footnotesize{}{}10.82}  & {\footnotesize{}{}(9.84, 11.27)}  & {\footnotesize{}{}10.15}  & {\footnotesize{}{}(9.57, 10.71)}  & {\footnotesize{}{}9.71}  & {\footnotesize{}{}(9.36,  10.16)}  \tabularnewline
{\footnotesize{}{}$\beta_{2}$}  & {\footnotesize{}{}1} &  {\footnotesize{}{}-}  & - & {\footnotesize{}{}-}  & {\footnotesize{}{}-}  & {\footnotesize{}{}0.87}  & {\footnotesize{}{}(0.39, 1.36)}  \tabularnewline
{\footnotesize{}{}$\sigma_{1}^{2}$}  & {\footnotesize{}{}4}&  {\footnotesize{}{}3.79}  & {\footnotesize{}{}(2.97, 5.19)} & {\footnotesize{}{}4.89}  & {\footnotesize{}{}(3.65, 6.27)}  & {\footnotesize{}{}3.51}  & {\footnotesize{}{}(2.71, 4.52)}   \tabularnewline
{\footnotesize{}{}$\sigma_{2}^{2}$}  & {\footnotesize{}{}1}&  {\footnotesize{}{}-}  & - & {\footnotesize{}{}-}  & {\footnotesize{}{}-}  & {\footnotesize{}{}1.05}  & {\footnotesize{}{}(0.18, 2.31)}    \tabularnewline
{\footnotesize{}{}$1/\phi_{1}$}  & {\footnotesize{}{}10} &  {\footnotesize{}{}13.29}  & {\footnotesize{}{}(9.33, 17.51)} & {\footnotesize{}{}8.75}  & {\footnotesize{}{}(6.49, 11.94)}  & {\footnotesize{}{}10.77}  & {\footnotesize{}{}(8.07, 13.91)}   \tabularnewline
{\footnotesize{}{}$1/\phi_{2}$}  & {\footnotesize{}{}10} &  {\footnotesize{}{}-}  & - & {\footnotesize{}{}-}  & {\footnotesize{}{}-}  & {\footnotesize{}{}12.61}  & {\footnotesize{}{}(3.93, 24.07)}    \tabularnewline
{\footnotesize{}{}$\gamma_{1}$}  & {\footnotesize{}{}1} &  {\footnotesize{}{}-}  & - & {\footnotesize{}{}-}  & {\footnotesize{}{}-}  & {\footnotesize{}{}0.995}  & {\footnotesize{}{}(0.983, 1.051)}  \tabularnewline
{\footnotesize{}{}$\tau_{1}^{2}$}  & {\footnotesize{}{}0.1}  &  {\footnotesize{}{}0.138}  & {\footnotesize{}{}(0.115, 0.183)}  &  {\footnotesize{}{}0.478}  & {\footnotesize{}{}(0.451, 0.508)}  & {\footnotesize{}{}0.125}  & {\footnotesize{}{}(0.097, 0.148)}    \tabularnewline
{\footnotesize{}{}$\tau_{2}^{2}$}  & {\footnotesize{}{}0.05} &  {\footnotesize{}{}-}  & -  & {\footnotesize{}{}-}  & {\footnotesize{}{}-}  & \footnotesize{}{}0.158{\footnotesize{} } & \footnotesize{}{}(0.041, 0.232){\footnotesize{} }   \tabularnewline
{\footnotesize{}{}$m$}  & {\footnotesize{}{}10}  &  {\footnotesize{}{}-}  & -  & {\footnotesize{}{}-}  & {\footnotesize{}{}-}  & {\footnotesize{}{}-}  & -\tabularnewline
\hline 
\end{tabular}{\footnotesize{}{}\caption{Unknown parameters are in the 1st column; their true values in the 2nd column;
their Bayesian point estimates and marginal credible intervals for
the single level NNGP, combined NNGP, and NNCGP models  are in the 3rd, 4th and 5th columns, respectively. The level 2 data set is used in the single level NNGP model, while both the level 1 and 2 data sets are used in the combined NNGP model and are treated as following a one level system. The estimated parameters by the single level NNGP and combined NNGP models are treated as level 1 parameters. \label{tab:Univariate-two-fildelity}}
} 
\end{table}

In Table \ref{tab:Univariate-two-fildelity-1}, we report standard
performance measures (defined in Appendix E) for the proposed NNCGP, single level NNGP, and combined NNGP models with $m=10$ neighbors. 
All performance measures indicate that NNCGP has better predictive ability than the single level NNGP and combined NNGP. 
NNCGP produces a significantly smaller effective number of model parameters (PD) and Deviance
Information Criterion (DIC) than the single level and combined NNGP, which suggests
that NNCGP provides a better fit when complexity is considered. The
root mean square prediction error (RMSPE) produced by NNCGP is approximately
$40$ - $50\%$ smaller than that of the single level NNGP and $20$ - $30\%$ smaller than that of the combined NNGP. The Nash-Sutcliffe model
efficiency coefficient (NSME) of NNCGP is closer to $1$ than that
of both other methods, which suggests that NNCGP provides a substantial improvement
in the prediction. 

\begin{table}
\center%
\begin{tabular}{c|ccc}
\hline 
\multirow{1}{*}{} & \multicolumn{3}{c}{{\footnotesize{}{}Model}}\tabularnewline
& {\footnotesize{}{}Single level NNGP}  & {\footnotesize{}{}Combined NNGP} & {\footnotesize{}{}NNCGP}\tabularnewline
\hline 
{\footnotesize{}{}RMSPE}  & {\footnotesize{}{}2.1325} &
{\footnotesize{}{}1.5202} &
{\footnotesize{}{}1.0987 }    \tabularnewline
{\footnotesize{}{}NSME }   &
{\footnotesize{}{}-1.1349 }  & {\footnotesize{}{}0.0888} & {\footnotesize{}{}0.5108 }  \tabularnewline
{\footnotesize{}{}CVG(95\%) }  & {\footnotesize{}{}0.8216} &
{\footnotesize{}{}0.7136 } &
{\footnotesize{}{}0.9573 }    \tabularnewline
{\footnotesize{}{}ALCI(95\%)}  & {\footnotesize{}{}5.2024} &
{\footnotesize{}{}3.0856} &
{\footnotesize{}{}3.2265 }   \tabularnewline
{\footnotesize{}{}PD }  & {\footnotesize{}{}12706} &
{\footnotesize{}{}5883} &
{\footnotesize{}{}2544 }   \tabularnewline
{\footnotesize{}{}DIC }  & {\footnotesize{}{}18136} &
{\footnotesize{}{}9659} &
{\footnotesize{}{}5551 }   \tabularnewline
\hline 
{\footnotesize{}{}Time(Hour) }  & {\footnotesize{}{}1.4} &
{\footnotesize{}{}2.9} &
{\footnotesize{}{}4.1 }   \tabularnewline
\hline 
\end{tabular}{\footnotesize{}{} \caption{Performance measures for the predictive ability of the NNCGP model, single level NNGP
model and combined NNGP model. (Definitions are given in Appendix E.) \label{tab:Univariate-two-fildelity-1}}
} 
\end{table}

In Figure \ref{fig:fig-4} we observe that, for the testing regions, the
NNCGP has more accurately captured the roughness and sharp changes
in the response surface while it also provides a better representation
of the patterns in the prediction surface. Applying NNGP directly to the
high-fidelity dataset provides a smoother prediction surface
due to the lack of the information from the low-fidelity dataset;
while it fails to produce reliable predictions in the blank regions. Applying NNGP to the combined dataset with both high- and low- fidelity levels (combined NNGP) also provides an unreliable prediction surface that is similar to the observations in low level regions. Modeling the scalar and  additive  discrepancies  between different levels helps improve predictions.
Moreover, NNCGP has produced a CVG closer to $0.95$ and a $95\%$
ALCI smaller than that of the single level NNGP and combined NNGP (Table \ref{tab:Univariate-two-fildelity}).
This indicates that NNCGP  produces more accurate predictions
with a higher probability to cover the true values in narrower credible
intervals.

\begin{figure}
\centering
\subfloat[{{{\large{}\label{fig:sfig2-1}}}}]{\includegraphics[width=0.495\linewidth]{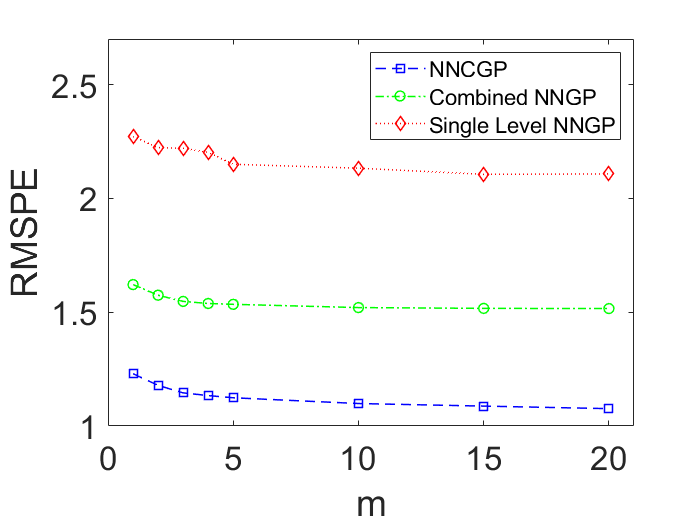}}{}{}
\subfloat[{{{\large{}\label{fig:sfig2-1}}}}]{{}{}\includegraphics[width=0.495\linewidth]{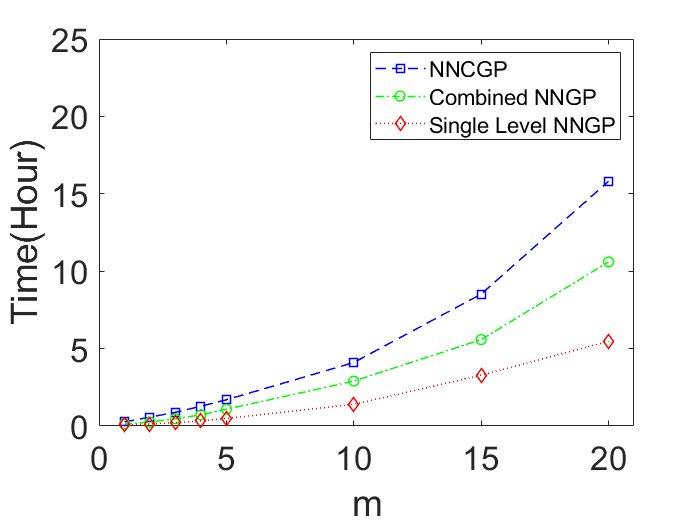}}
\caption{{\large{}{}{}{}{}{}{}\label{fig:fig-sensitive} } Sensitivity analysis to number of nearest neighbors $m$: a) Root mean square prediction error (RMSPE) and b) Running time for the NNCGP, single level NNGP, and combined NNGP models for a range of $m$ values over a two fidelity level non-nested synthetic dataset.}
\end{figure}

To test the sensitivity of the proposed NNCGP method to the number of neighbors $m$, we compare the  RMSPEs of the three different methods for $m = \{1,2,3,4, 5, 10, 15\}$, and $20$ in Figure~\ref{fig:fig-sensitive}(a). We use the same prior specifications and  computational strategies as described above. In terms of  prediction accuracy, the NNCGP outperforms both the single level NNGP and the combined NNGP for all $m$. The decrease of the RMSPE is smaller as $m$ becomes greater than $10$. The computational time is longer for the NNCGP  than both the single level NNGP and combined NNGP (Figure~\ref{fig:fig-sensitive}(b)).
The NNCGP  uses data from both levels and also expands the reference set to ensure the nested structure between different levels.  The single NNGP model only uses data from the high-fidelity level. The combined NNGP uses the same dataset as the NNCGP. However, the reference set of the combined NNGP is equal to the reference set of the first fidelity level of NNCGP. It is worth mentioning that  if  the observation locations are nested or semi-nested the computational complexity of the NNCGP can be further reduced since $\tilde{n}_1<(n_1+n_2)$.


\section{Application to intersatellite calibration}

Satellite soundings have been providing measurements of the Earth’s
atmosphere, oceans, land, and ice since the 1970s to support the study
of global climate system dynamics. Long term observations from past
and current environmental satellites are widely used in developing
climate data records (CDR) \citep{nrc2004}. We examine here one instrument
in particular, the high-resolution infrared radiation sounder (HIRS)
instrument that has been taking measurements since 1978 on board the
National Oceanic and Atmospheric Administration (NOAA) polar orbiting
satellite series (POES) and the meteorological operational satellite
program (Metop) series operated by the European Organization for the
Exploitation of Meteorological Satellites (EUMETSAT). This series
of more than a dozen satellites currently constitutes over 40 years
of HIRS observations, and this unique longevity is valuable to characterize
climatological trends. Examples of essential climate variables derived
from HIRS measurements include long-term records of temperature and
humidity profiles \citep{shi2016,matthews2019}.

HIRS mission objectives include observations of atmospheric temperature,
water vapor, specific humidity, sea surface temperature, cloud cover,
and total column ozone. The HIRS instrument is comprised of twenty
channels, including twelve longwave channels, seven shortwave channels,
and one visible channel. Among the longwave channels, Channels 1 to
7 are in the carbon dioxide (CO2) absorption band to measure atmospheric
temperatures from near-surface to stratosphere, Channel 8 is a window
channel for surface temperature observation and cloud detection, Channel
9 is an ozone channel, and Channels 10--12 are for water vapor signals
at the near-surface, mid-troposphere, and upper troposphere, respectively.
There have been several versions of the instruments where there is
a notable change in spatial resolution. In particular, for the HIRS/2
instrument, with observations from the late 1970s to mid-2007, the
spatial footprint is approximately 20 km. HIRS/3, with observations
from 1998 to mid-2014, has a spatial footprint of approximately 18
km. The currently operational version, HIRS/4, improved the spatial
resolution to approximately 10 km at nadir with observations beginning
in 2005. The dataset being considered in this study is limb-corrected
HIRS swath data as brightness temperatures \citep{jackson2003}. The
data is stored as daily files, where each daily file records approximately
120,000 geolocated observations. The current archive includes data
from NOAA-5 through NOAA-17 along with Metop-02, covering the time
period of 1978-2017. In all, this data archive is more than 2 TB,
with an average daily file size of about 82 MB.

The HIRS data record faces some common challenges when developing
CDRs from the time series. Specifically, there are consistency and
accuracy issues due to degradation of sensors and intersatellite discrepancies.
Furthermore, there is missing information caused by atmospheric
conditions such as thick cloud cover. As early as 1991, to address
some of these challenges, the co-kriging technique has been applied
to remotely sensed data sets \citep{bhatti1991}. As an improvement
to these techniques, we consider using the NNCGP model as a method
for intersatellite calibration, data imputation, and data prediction.

\begin{figure}
{}{}\subfloat[{{{\large{}\label{fig:sfig1-1} }Observations
of NOAA 14}}]{{}{}\includegraphics[width=0.49\linewidth]{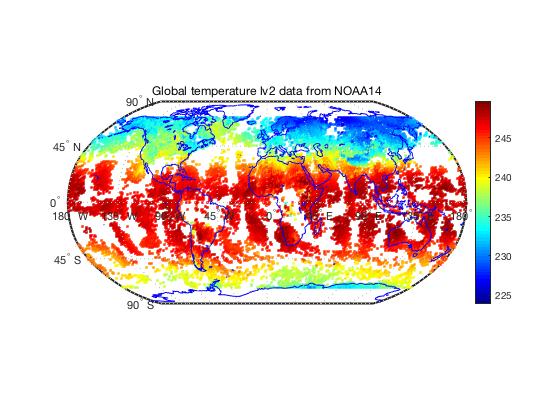}

}{}{}\subfloat[{{{\large{}\label{fig:sfig2-1}}Training
data of NOAA 15}}]{{}{}\includegraphics[width=0.49\linewidth]{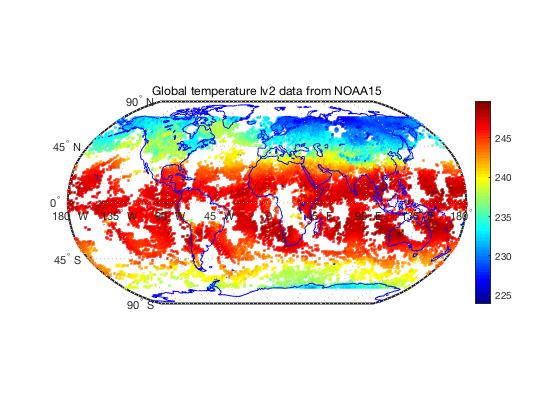}
}{}{}
\caption{{\large{}{}{}{}{}{}{}\label{fig:fig-1} } Brightness
temperature observations by HIRS Channel 5 on March 1, 2001 by (a) NOAA-14 and (b) NOAA-15.}
\end{figure}

We examine HIRS Channel 5 observations from a single day, March
1, 2001, as illustrated in Figure \ref{fig:fig-1}. On this day, we
may exploit a period of temporal overlap in the NOAA POES series where
two satellites captured measurements: NOAA-14 and NOAA-15. The HIRS
sensors on these two satellites have similar technical designs which
allow us to ignore the spectral and spatial footprint differences.
NOAA-14 became operational in December 1994 while NOAA-15 became operational
in October 1998. Given the sensor age difference, it is reasonable
to consider that the instruments on-board NOAA-15 are in better condition
than those of NOAA-14. Therefore, we treat observations from NOAA-14
as a dataset of low fidelity level, and those from NOAA-15 as a dataset
of high fidelity level. A small region of observations from NOAA-15
are treated as testing data, and the remainder of the NOAA-15 observations are treated as
training data.

We model our data based on the two-fidelity level NNCGP model as described in Sections 3 \& 4.  We consider a linear model for the mean  of the Gaussian processes in $y_1(\cdot)$ and $\delta_{2}(\cdot)$  with a linear basis function representation  $\{\mathbf{h}(s_{t})\}$
and coefficients $\boldsymbol{\beta}_{t}=\{\beta_{0,t},\beta_{1,t},\beta_{2,t}\}^{T}$. We consider the scalar discrepancy $\zeta(s)$ to be an unknown constant  and equal to $\gamma$. The number of nearest neighbors $m$ is set to 10, and the spatial process $\bfw_{t}$ is considered  to have a diagonal anisotropic exponential covariance function as described in Section 3. 

We assign independent normally distributed priors with zero mean and large  variances for $\beta_{0,t},\beta_{1,t},\beta_{2,t}$ and 
$\gamma$.  We assign independent uniform prior distributions $U(0,1000)$ to the range correlation parameters $({\phi}_{t,1},{\phi}_{t,2})$ for $t=1,2$. Also, we assign independent $IG(2,1)$ prior distributions for the variance parameters $\sigma_{t}^{2}$ and $\tau_{t}^{2}$. For the Bayesian inference of the NNCGP, we run the MCMC sampler described in Section $4$ with  $35,000$ iterations where the first $5,000$ iterations are discarded as a burn-in. 

\begin{table}[H]
{}{}\centering %
\begin{tabular}{c|ccc}
\hline 
\multirow{1}{*}{{}{} } & \multicolumn{3}{c}{{}{}Model}\tabularnewline
 & {}{}NNCGP  & {}{}Single level NNGP & {}{}Combined NNGP \tabularnewline
\hline 
{}{}RMSPE  & {}{}1.2044  & {}{}1.8153 & {}{}1.6772 \tabularnewline
{}{}NSME  & {}{}0.8439  & {}{}0.5499 & {}{}0.6726 \tabularnewline
{}{}CVG(95\%)  & {}{}0.9255  & {}{}0.8350 & {}{}0.9197 \tabularnewline
{}{}ALCI(95\%)  & {}{}3.094  & {}{}4.214 & {}{}5.778 \tabularnewline
\hline 
{}{}Time(Hour)  & {}{}38  & {}{}20 & {}{}32 \tabularnewline
\hline 
\end{tabular}{}{}\caption{{\large{}{}{}{}{}{}{}\label{real_data_table}}Performance measures for the predictive ability of the NNCGP model, single level NNGP
model and combined NNGP models for Channel 5 NOAA-14 and NOAA-15 observations on March 1, 2001. (Definitions are given in Appendix E.)}
\end{table}

\begin{figure}
{}{}\subfloat[{{{\large{}\label{fig:sfig1-3}}NOAA-15 testing data-set}}]{{}{}\includegraphics[width=0.45\linewidth]{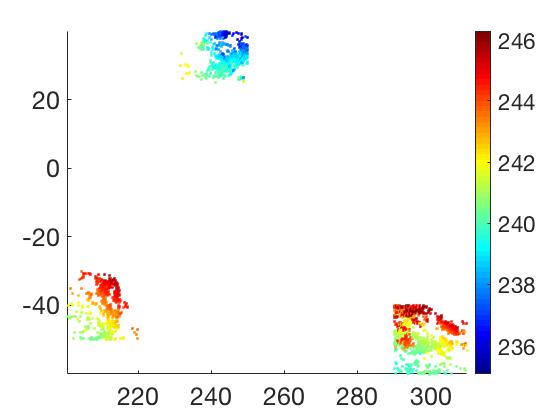}
}{}{}\hspace{0.5cm}
\subfloat[{{{\large{}\label{fig:sfig1-2}}Prediction means by NNCGP model}}]{{}{}\includegraphics[width=0.45\linewidth]{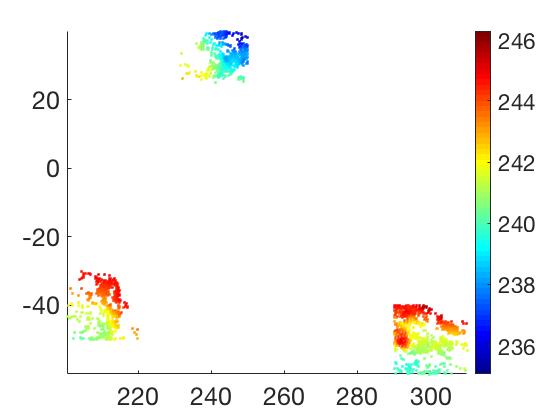}
}{}{} \hfill
\subfloat[{{{\label{fig:sfig2-2} }Prediction means by single level NNGP
model}}]{{}{}\includegraphics[width=0.45\linewidth]{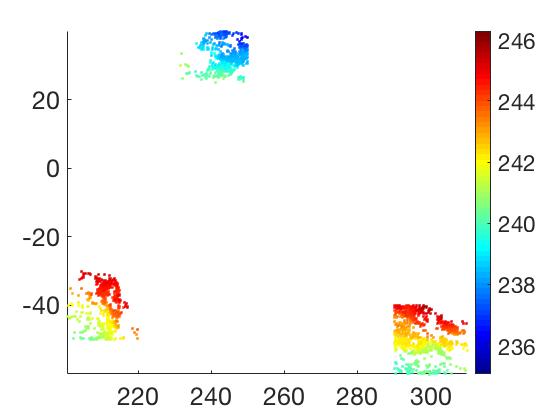}
}{}{}\hspace{1.2cm}
\subfloat[{{{\label{fig:sfig2-2} }Prediction means by combined NNGP
model}}]{{}{}\includegraphics[width=0.45\linewidth]{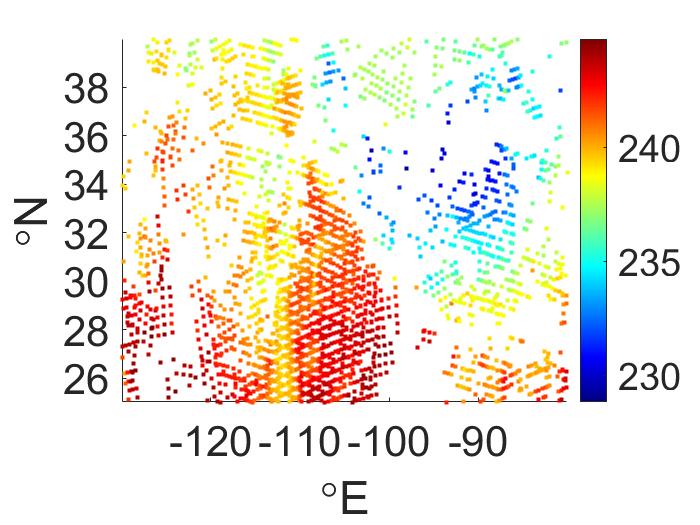}
}{}{}
\caption{{\large{}{}{}{}{}{}{}\label{fig:fig-2} }Predictions of NOAA-15
Brightness Temperatures(K) testing data-set by NNCGP, single level
NNGP and combined NNGP under fully non-nested experimental design.}
\end{figure}

The prediction performance metrics of the three different methods are
given in Table \ref{real_data_table}.   Compared to the single level NNGP and combined NNGP models, the NNCGP model
produced an approximately 30\% smaller RMSPE and its NSME is closer to 1. The NNCGP also produced a larger CVG and a smaller ALCI than the single level and combined NNGP models. The results suggest that the NNCGP model provides a substantial improvement
in terms of predictive accuracy in real data analysis. The NOAA-15 testing data shown in Figure
\ref{fig:fig-2} shows that the NNCGP model is more capable of capturing the spatial patterns of the
testing data than either a single level or a combined NNGP model. Unlike the single level NNGP,  the proposed NNCGP uses  additional information from NOAA-14.  Compared to a combined NNGP, the proposed NNGCP benefits from modeling the discrepancy of observations from the different satellites. With the fully non-nested structure, the computational complexity
of the single level NNGP model is $\mathcal{O}(n_{2}m^{3})$ and
for the NNCGP model is $\mathcal{O}((n_{1}+n_{2})m^{3})$; this is consistent with the running times of the models shown in Table \ref{real_data_table}. 

Baseline observations  of brightness temperature  are used as inputs for so-called remote sensing retrieval algorithms wherein thematic climate variables (e.g. precipitation rates, cloud cover, surface temperature, etc.) are derived. These retrieval algorithms are typically highly nonlinear, so a small change in the input brightness temperature value can have a large impact on the value of derived climate variables. The significant improvement in the prediction accuracy to the brightness temperatures provided by the NNCGP model is therefore critical to the downstream climate variables because of this sensitivity. So although the computation by the NNCGP model is costly compared to the single level NNGP model, it is still worthwhile to  apply the NNCGP model.

\begin{figure}
{}{}\centering \includegraphics[width=0.9\linewidth]{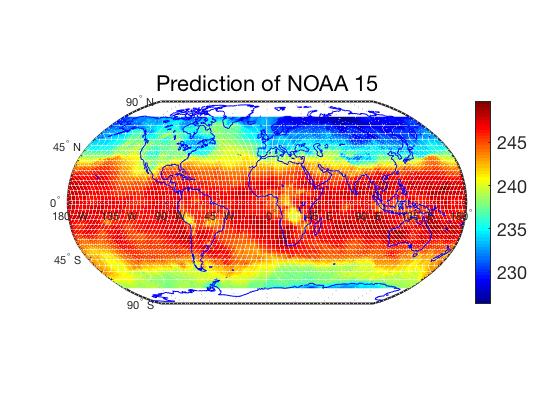}
\caption{{\large{}{}{}{}{}{}{}\label{fig:sfig2-2-1}}Gap-filled global brightness temperature predictions
of NOAA-15 data on a regular grid.}
\label{fig:fig-7} 
\end{figure}

We applied the NNCGP model to gap-fill predictions
based upon a discrete global grid. We chose to use $1^{\circ}$ latitude
by $1.25^{\circ}$ longitude ($1^{\circ}\times1.25^{\circ}$) pixels
as a grid structure with near-global spatial coverage from $-70^{\circ}$ to $70^{\circ}$N.
By applying the NNCGP model, we predict gridded NOAA-15 brightness
temperature data on the center of the grids, based on the NOAA-14
and NOAA-15 swath-based spatial supports. The prediction plot (Figure
\ref{fig:fig-7}) illustrates the ability of the NNCGP model to handle
large, irregularly spaced data sets and produce a gap-filled composite
gridded data-set.

\section{Summary and conclusions}

In this manuscript we have proposed a new, computationally efficient Nearest Neighbor Autoregressive
Co-Kriging Gaussian process (NNCGP) method for the analysis of large irregularly spaced and
multi-fidelity spatial data. The proposed NNCGP method extends the
scope of the classical auto-regressive co-kriging models  \citep{kennedy2000predicting,konomikaragiannisABTCK2019}
to deal with large data sets. To deal with this complexity, we used independent NNGP priors at each level in the  auto-regressive co-kriging model where the neighboors are simply defined within each level. However,  when the observed locations are hierarchically non-nested, the likelihood does not simplify; this makes the computational complexity of this approach infeasible. To overcome this issue, we augment the spatial random effects such that they can form a suitable nested structure. The augmentation of the spatial random effects facilitates the computation of  the likelihood of NNCGP. This is because it enables the factorization of the likelihood into terms with smaller covariance matrices and gains the computational efficiency provided by the nearest neighbors 
within each level. The proposed method is  at most computationally linear in the total number of all spatial locations of all fidelity levels. 
In our simulations, we observed that computations were faster when the observed locations  at higher fidelity were nested within those at lower fidelity levels. Moreover, the nested design of the reference sets allows
the assignment of semi-conjugate priors for the majority of the
parameters. Based on these specifications, we develop efficient and independent MCMC block updates for Bayesian inference. As in the original NNGP paper \citep{datta2016hierarchical}, our results indicate that inference is very robust with respect to values of neighbors.

We compared the proposed NNCGP with  NNGP in the single level of highest-fidelity
with a simulation study and a real data application of intersatellite
calibration. We observed that the NNCGP was able to improve the prediction accuracy
of HIRS brightness temperatures from the NOAA-15 polar-orbiting
satellite by incorporating information from an older version of the
same HIRS sensor onboard the polar-orbiting satellite NOAA-14. Beyond HIRS, the
proposed methodology can be used for a variety of large multi-fidelity
level remotely sensed data sets  with  overlapping observations from sensors of similar design. Furthermore, we propose that the proposed methodology can be used in a wide
range of applications in physical science and engineering when multiple computer models with large simulation runs are available.

Several extensions of the proposed NNCGP method can be pursued in future work.   Based on  the resulting conditional posterior distributions,  the  Bayesian inference can be derived in parallel for each fidelity level similarly to how is done for the NNGP \citep{datta2016hierarchical}. In this case, if parallel computing is available, the computational complexity of the approach can be reduced up to $\mathcal{O}(\tilde{n}_{1}m^{3})$.  Another  possible extension is to accelerate our inferential method with a reference or conjugate NNCGP based on ideas given in \citet{finley2019efficient}.  The proposed model may also be extended in the multivariate setting by using  parallel partial autoregressive co-kriging \citep{ma2019multifidelity}. Finally, the proposed NNCGP model can be extended to spatial-temporal settings with discretized time steps, as both autoregressive structures and the NNGP approach are capable of incorporating temporal dependence.

\section*{Acknowledgments}

Matthews was supported by National Oceanic and Atmospheric Administration (NOAA) through the Cooperative Institute for Climate and Satellites - North Carolina under Cooperative Agreement NA14NES432003 and the Cooperative Institute for Satellite Earth System Studies under Cooperative Agreement NA19NES4320002.

\singlespacing \setlength{\bibsep}{5pt}

 \bibliographystyle{jasa3}
\bibliography{reference1}

\appendix

\section*{Appendix}

\section{NNGP specifications}\label{App1}

The posterior distribution of {\small{}{}{}
\begin{align}
\tilde{p}(\bfw_{t}|\cdot) & \propto\text{exp}\left[-\frac{1}{2}\sum_{i=1}^{n_{t}}\left\{ w_{t}(s_{t,i})-\mathbf{B}_{t,s_{t,i}}\bfw_{t,N_{t}(s_{t,i})}\right\} ^{T}F_{t,s_{t,i}}^{-1}\left\{ w_{t}(s_{t,i})-\mathbf{B}_{t,s_{t,i}}\bfw_{t,N_{t}(s_{t,i})}\right\} \right]\nonumber \\
 & =\text{exp}\left(-\frac{1}{2}\bfw_{t}^{T}\mathbf{B}_{t}^{T}\mathbf{F}_{t}^{-1}\mathbf{B}_{t}\bfw_{t}\right),
\end{align}
}{\small\par}

where $\mathbf{F}_{t}=\text{diag}(F_{t,s_{t,1}},F_{t,s_{t,2}},\ldots,F_{t,s_{t,n_{t}}})$,
$\mathbf{B}_{t}=\Big{(}\mathbf{B}_{t,1}^{T},\mathbf{B}_{t,2}^{T},\ldots,\mathbf{B}_{t,n_{t}}^{T}\Big{)}^{T}$,
and for each element in $\mathbf{B}_{t}$, we have $\mathbf{B}_{t,i}=\Big{(}\mathbf{B}_{t,s_{t,i},1}^{T},\mathbf{B}_{t,s_{t,i},2}^{T},\ldots,\mathbf{B}_{t,s_{t,i},n_{t}}^{T}\Big{)}^{T}$
and {\small{}{}{} 
\begin{align}
\mathbf{B}_{t,s_{t,i},j}=\begin{cases}
1,\ \text{if}\ i=j,\\
-\mathbf{B}_{t,s_{t,i}}[,k],\ \text{if}\ s_{t,j}\ \text{is the}\ k^{th}\ \text{element in}\ N_{t}(s_{t,i}),\\
0,\ \text{Others}.
\end{cases}
\end{align}
}

\section{Mean and Variance Specifications}

The mean vector $\boldsymbol{\mu} =  \left(\mu_1(\bs_{1,1}),\ldots,\mu_1(\bs_{1,n_1}),\ldots,\mu_T(s_{T,n_T})\right)$ is
\allowdisplaybreaks
\begin{align}
\mu_t(\bs_{t,k})=&\mathbf{1}_{\{t>1\}}(t)\sum_{i=1}^{t-1}\left\{\prod_{j=i}^{t-1}\zeta_{j}(\bs_{t,k})\right\} \left\{\mathbf{h}_{i}^{T}(\bs_{t,k})\boldsymbol{\beta}_{i} + \mathbf{1}_{\{\bs_{t,k}\in \bfS_i\}}(\bs_{t,k})w_i(\bs_{t,k}) \right\} \nonumber \\ &+\mathbf{h}_{t}^{T}(s_{t,k})\boldsymbol{\beta}_{t} + w_t(\bs_{t,k}),  \label{mean_function_original}
\end{align}
for $t=1,\ldots T$, $  i=1,\ldots, n_t$. $\mathbf{1}_{\{\cdot\}}(\cdot)$ is the indicator function,  and covariance matrix $\boldsymbol{\Lambda}$ is a block matrix with blocks $\Lambda^{(1,1)},\ldots,\Lambda^{(1,T)},\ldots,\Lambda^{(T,T)}$, and the size of $\boldsymbol{\Lambda}$ is $\sum_{t=1}^{T}n_t \times \sum_{t=1}^{T}n_t$. The $\Lambda^{(t,t)}$ 
components are calculated as:
\begin{align}
 & \Lambda^{(t,t)}_{k,l} =  \text{cov}(z_{t}(\bs_{t,k}),z_{t}(\bs_{t,l})|\cdot)=\sum_{i=1}^{t-1} \mathbf{1}_{\{\bs_{t,k},\bs_{t,l}\notin \bfS_i\}}(\bs_{t,k},\bs_{t,l}) \left\{ \prod_{j=i}^{t-1}\zeta_{j}(\bs_{t,k})^{T}\zeta_{j}(\bs_{t,l})\right\} C_{i}(\bs_{t,k},\bs_{t,l}|\boldsymbol{\theta}_{i}) \nonumber \\
 & \qquad \qquad \qquad  +\mathbf{1}_{\bs_{t,k}=\bs_{t,l}}(\bs_{t,k},\bs_{t,l})\tau_{t}^{2},\nonumber 
 \end{align}
for $t \ \text{and}\ t' = 1,\ldots,T$; $k=1,\ldots,n_t$; $l = 1,\ldots,n_{t'}$,  and
 \begin{align}
 &\Lambda^{(t,t')}_{k,l} =  \text{cov}(z_{t}(\bs_{t,k}),z_{t'}(\bs_{t',l})|\cdot)=\sum_{i=1}^{\text{min}(t,t')-1} \mathbf{1}_{\{\bs_{t,k},\bs_{t',l}\notin \bfS_i\}}(\bs_{t,k},\bs_{t',l}) \left\{ \prod_{j=i}^{\text{min}(t,t')-1}\zeta_{j}(\bs_{t,k})^{T}\zeta_{j}(\bs_{t',l})\right\} \nonumber \\ 
 & \qquad \qquad \qquad  \times C_{i}(\bs_{t,k},\bs_{t',l}|\boldsymbol{\theta}_{i})  + \mathbf{1}_{\{\bs_{t,k},\bs_{t',l}\notin \bfS_{\text{min}(t,t')}\}}(\bs_{t,k},\bs_{t',l}) C_{\text{min}(t,t')}(\bs_{t,k},\bs_{t',l}|\boldsymbol{\theta}_{\text{min}(t,t')}),\label{covariance_function_original}
\end{align}
for $t\neq t'$, $\Lambda^{(t,t')}$. 

\section{Alternative form of joint posterior distribution}
The joint posterior density function of NNCGP for a $T$ level system is:
\begin{align}
p(\boldsymbol{\Theta}_{1:T},\tilde{\bfw}_{1:T}|\bfZ_{1:T})= & \prod_{t=1}^{T}\Bigg{[}p(\boldsymbol{\Theta}_{t})\tilde{p}(\bfw_{t}|\boldsymbol{\theta}_{t})\tilde{p}(\bfw_{t}^{*}|\bfw_{t},\boldsymbol{\theta}_{t})\Bigg{]} L(\boldsymbol{Z}_{1:T}|\boldsymbol{\beta}_{1:T},\tau_{1:T}^{2},\bfgamma_{1:T-1},\tilde{\bfw}_{1:T}),
\end{align}

The conditional joint likelihood is 
\begin{align}
 & L(\boldsymbol{Z}_{1:T}|\boldsymbol{\beta}_{1:T},\tau_{1:T}^{2},\bfgamma_{1:T-1},\tilde{\bfw}_{1:T})=\prod_{t=1}^{T}\prod_{i=1}^{n_{t}}N\bigg{(}z_{t}(\bs_{t,i})|\mu_{t}(\bs_{t,i}),\tau_{t}\bigg{)},\nonumber \\
 & \mu_{t}(\bs_{t,i})=\mathbbm{1}_{\{t\not=1\}}(t)\times\sum_{j=1}^{t-1}\Bigg{\{}[\prod_{k=j}^{t-1}\zeta_{k}(\bs_{t,i})]\times[\mathbf{h}_{j}(\bs_{t,i})^{T}\boldsymbol{\beta}_{j}+w_{j}(\bs_{t,i})]\Bigg{\}}+\mathbf{h}_{t}(\bs_{t,i})^{T}\boldsymbol{\beta}_{t}+w_{t}(\bs_{t,i}).
\end{align}

We assume $\bfw_{t}^{*}|\bfw_{t}$
are independent from each other so they can  be updated individually.   For locations $\bs_{u}\in\bfS_{t}^{*}$, the full conditional posterior distribution of $w_{t}(\bs_{u}) \sim N(V_{t}^*(\bs_u)\mu_{t}^*(\bs_u),V_{t}^*(\bs_u))$ with 
\begin{align}
V_{t}^*(\bs_u)= & \Bigg{[}\sum_{q=t+1}^{T}(\prod_{i=t}^{q-1}I_{\{\boldsymbol{U}_{t,q}\}}(\bs_u)\zeta_{i}^{2}(\bs_u)\tau_{q}^{-2})+F_{t,\bs_u}^{-1}\Bigg{]}^{-1},\nonumber \\
\mu_{t}^*(\bs_u)= & \sum_{q=t+1}^{T}\prod_{i=t}^{q-1}I_{\{\boldsymbol{U}_{t,q}\}}(\bs_u)\zeta_{i}(\bs_u)\tau_{q}^{-2}\Bigg{(}z_{q}(\bs_u)-[\zeta_{q-1}(\bs_u)y_{q-1}(\bs_u)-\prod_{i=t}^{q-1}\zeta_{i}(\bs_u)w_{t}(\bs_u)] \Bigg{)}\nonumber \\
 & +F_{t,\bs_u}^{-1}\mathbf{B}_{t,\bs_u}\bfw_{t,N_{t}(\bs_u)},\ \ t=1,2,\ldots,T-1, 
 \label{imputation_w}
\end{align}
where $I_{\{\boldsymbol{U}\}}(\bs_u)$ is an indicator function with value $1$ for location $\bs_u \in \boldsymbol{U}$ and 0 for $\bs_u \notin \boldsymbol{U}$, $\boldsymbol{U}_{A,B} = \bfS_A \cap \bfS_B$.

\section{Gibbs Sampler }

The full conditional distribution of  $\bfw_{1}$ is  
\begin{align}
w_{1}(\bs_{u})|\boldsymbol{\Theta}_{1},\bfZ_1 & \sim N(V_{\bfw_{1}}(\bs_{u})\mu_{\bfw_{1}}(\bs_{u}),V_{\bfw_{1}}(\bs_{u})),\nonumber \\
V_{\bfw_{1}}(\bs_{u}) & =  \left(F_{1,\bs_{u}}^{-1}+\tau_{1}^{-2}\right)^{-1},\nonumber \\
\mu_{\bfw_{1}}(\bs_{u}) & =  \tau_{1}^{-2}[z_{1}(\bs_{u})-\mathbf{h}_{1}^T(\bs_{u})\boldsymbol{\beta}_{1}]+F_{1,\bs_{u}}^{-1}\mathbf{B}_{1,\bs_{u}}\bfw_{1,N_{1}(\bs_{u})},\label{w_sampler}
\end{align}
for $s_{u}\in \bfS_{1}$.

With the specification of priors, the posterior distributions of the parameters are: {\small{}{}{} 
\begin{align}
  \boldsymbol{\beta}_{t}|\bfw_{t},\tilde{\bfy}_{t-1},\bfgamma_{t-1},\tau_{t},\bfZ_{t} &\sim N(\mathbf{V}_{\boldsymbol{\beta}_{t}}^{*}\boldsymbol{\mu}_{\boldsymbol{\beta}_{t}}^{*},\mathbf{V}_{\boldsymbol{\beta}_{t}}^{*}),\nonumber \\
  \bfgamma_{t}|\tilde{\bfy}_{t},\boldsymbol{\beta}_{t+1},\tau_{t+1},\bfZ_{t+1} & \sim N(\mathbf{V}_{\bfgamma_{t}}^{*}\boldsymbol{\mu}_{\bfgamma_{t}}^{*},\mathbf{V}_{\bfgamma_{t}}^{*}),\nonumber \\
  \sigma_{t}^{2}|\tilde{\bfw}_{t},\boldsymbol{\phi}_{t} & \sim IG(a_{\sigma_{t}}^{*},b_{\sigma_{t}}^{*}),\nonumber \\
 \tau_{t}^2|\boldsymbol{\beta}_{t},\bfw_{t},\tilde{\bfy}_{t-1},\bfZ_{t},\bfgamma_{t-1} & \sim IG(a_{\tau_{t}},b_{\tau_{t}}),\label{gibbs_sampler}
\end{align}
}For $\boldsymbol{\beta}_{t}$, we have: 
\begin{align*}
p(\boldsymbol{\beta}_{t}|\cdot) & \propto N(\boldsymbol{\beta}_{t}|\boldsymbol{\mu}_{\boldsymbol{\beta}_{t}},\mathbf{V}_{\boldsymbol{\beta}_{t}})\times N(\bfZ_{t}|\mathbf{1}_{t>1}(t)\zeta_{t-1}\circ y_{t-1}(\bfS_t)+\boldsymbol{\delta}_{t},\tau_{t}\mathbf{I})  \\
 & \propto\text{exp}\bigg\{-\frac{1}{2}(\boldsymbol{\beta}_{t}-\boldsymbol{\mu}_{\boldsymbol{\beta}_{t}})^{T}\mathbf{V}_{\boldsymbol{\beta}_{t}}^{-1}(\boldsymbol{\beta}_{t}-\boldsymbol{\mu}_{\boldsymbol{\beta}_{t}})\bigg\}\times\\
 & \quad\;\text{exp}\bigg[-\frac{1}{2\tau_{t}^{2}}\bigg\{ \bfZ_{t}-\mathbf{1}_{t>1}(t)\zeta_{t-1}\circ y_{t-1}(\bfS_t)-\boldsymbol{\delta}_{t}\bigg\}^{T}\bigg\{ \bfZ_{t}-\mathbf{1}_{t>1}(t)\zeta_{t-1}\circ y_{t-1}(\bfS_t)-\boldsymbol{\delta}_{t}\bigg\}\bigg]\\
 & \propto\text{exp}\bigg{\{}-\frac{1}{2}\bigg\{\mathbf{V}_{\boldsymbol{\beta}_{t}}^{-1}+\frac{1}{\tau_{t}^{2}}\mathbf{h}_{t}\mathbf{h}_{t}^{T}\bigg\}\boldsymbol{\beta}_{t}^{T}\boldsymbol{\beta}_{t}+\\
 & \quad\;\left[\boldsymbol{\mu}_{\boldsymbol{\beta}_{t}}^{T}\mathbf{V}_{\boldsymbol{\beta}_{t}}^{-1}+\frac{1}{\tau_{t}^{2}}(\bfZ_{t}-\mathbf{1}_{t>1}(t)\zeta_{t-1}\circ y_{t-1}(\bfS_t)-\bfw_{t})^{T}\mathbf{h}_{t}^T\right]\boldsymbol{\beta}_{t}\bigg{\}},
\end{align*}
and we have: 
\begin{align}
 \boldsymbol{\beta}_{t}|\bfw_{t},\tilde{\bfy}_{t-1},\bfgamma_{t-1},\tau_{t},\bfZ_{t} & \sim N(\mathbf{V}_{\boldsymbol{\beta}_{t}}^{*}\boldsymbol{\mu}_{\boldsymbol{\beta}_{t}}^{*},\mathbf{V}_{\boldsymbol{\beta}_{t}}^{*}),\nonumber \\
  \boldsymbol{\mu}_{\boldsymbol{\beta}_{t}}^{*} & =\boldsymbol{\mu}_{\boldsymbol{\beta}_{t}}^{T}\mathbf{V}_{\boldsymbol{\beta}_{t}}^{-1}+\frac{1}{\tau_{t}^{2}}(\bfZ_{t}-\mathbf{1}_{t>1}(t)\zeta_{t-1}\circ y_{t-1}(\bfS_t)-\bfw_{t})^{T}\mathbf{h}_{t}^T,\nonumber \\
 \mathbf{V}_{\boldsymbol{\beta}_{t}}^{*} & =\left( \mathbf{V}_{\boldsymbol{\beta}_{t}}^{-1}+\frac{1}{\tau_{t}^{2}}\mathbf{h}_{t}\mathbf{h}_{t}^{T}\right) ^{-1}, \label{eq.beta_t}
\end{align}
where $\mathbf{1}_{t>1}(t)$ is an indicator function equals 1 for $t>1$,
otherwise 0. The full conditional distribution for parameter $\sigma_{t}^{2}$
is: 
\begin{align*}
p(\sigma_{t}^{2}|a_{t},b_{t},w_{t}(\mathbf{\tilde{S}}_{t}),\boldsymbol{\phi}_{t}) & \propto IG(a_{t},b_{t})\times\tilde{p}(\mathbf{w}_{t})\times\tilde{p}(w_{t}(\mathbf{S}_{t}^{*})|\mathbf{w}_{t}).
\end{align*}
From nearest neighbor Gaussian process approach, we have 
\begin{align*}
  \tilde{p}(\mathbf{w}_{t})= & \prod_{i=1}^{n_{t}}p(w_{t}(\bs_{t,i})|\bfw_{t,N_{t}\bs_{t,i})}),\\
  w_{t}(\bs_{t,i})|\bfw_{t,N_{t}(\bs_{t,i})})\sim & N(\mathbf{B}_{\bs_{t,i}}\bfw_{t,N_{t}(\bs_{t,i})},F_{\bs_{t,i}}),
\end{align*}
where $\mathbf{B}_{s_{t,i}}=\mathbf{C}_{\bs_{t,i},N(\bs_{t,i})}^T\mathbf{C}_{N(\bs_{t,i})}^{-1}$,
$\mathbf{F}_{\bs_{t,i}}=\mathbf{C}_{\bs_{t,i},s_{t,i}}-\mathbf{C}_{\bs_{t,i},N(\bs_{t,i})}^T\mathbf{C}_{N(\bs_{t,i})}^{-1}\mathbf{C}_{\bs_{t,i},N(\bs_{t,i})}$,
here $\mathbf{C}$ is the covariance matrix. Denote $\mathbf{C}_{t}(.,.|\sigma_{t}^{2},\phi_{t})=\sigma_{t}^{2}\boldsymbol{\rho}_{t}(.,.|\phi_{t})$,
we have: 
\begin{align*}
\mathbf{F}_{\bs_{t,i}}=\sigma_{t}^{2}(\boldsymbol{\rho}(\bs_{t,i},\bs_{t,i})-\boldsymbol{\rho}_{\bs_{t,i},N(\bs_{t,i})}\boldsymbol{\rho}_{N(\bs_{t,i})}^{-1}\boldsymbol{\rho}_{\bs_{t,i},N(\bs_{t,i})})=\sigma_{t}^{2}\tilde{\mathbf{F}}_{\bs_{t,i}},
\end{align*}
and the following full conditional distribution for $\sigma_{t}^{2}$
is: 
\begin{align*}
 & p(\sigma_{t}^{2}|a_{t},b_{t},\mathbf{\tilde{w}}_{t},\boldsymbol{\phi}_{t})\propto(\sigma_{t}^{2})^{-a_{t}-1}\text{exp}(-\frac{b_{t}}{\sigma_{t}^{2}})\times\tilde{p}(\mathbf{w}_{t})\times\tilde{p}(w_{t}(\mathbf{S}_{t}^{*})|\mathbf{w}_{t})\\
 & \propto(\sigma_{t}^{2})^{-a_{t}-1-\frac{1}{2}(n_{t}+n_{t}^{*})}\text{exp}\bigg[-\frac{1}{2\sigma_{t}^{2}}\sum_{i=1}^{n_{t}}\left\{ w_{t}(\bs_{t,i})-\mathbf{B}_{\bs_{t,i}}\bfw_{t,N_{t}(s_{t,i})}\right\} ^{T}(\tilde{F}_{\bs_{t,i}})^{-1}\left\{ w_{t}(s_{t,i})-\mathbf{B}_{\bs_{t,i}}\bfw_{t,N_{t}(\bs_{t,i})}\right\} \\
 & -\frac{1}{2\sigma_{t}^{2}}\sum_{i=1}^{n_{t}^{*}}\left\{ w_{t}(s_{t,i}^{*})-\mathbf{B}_{\bs_{t,i}^{*}}\bfw_{t,N_{t}(\bs_{t,i}^{*})}\right\} ^{T}(\tilde{F}_{\bs_{t,i}^{*}})^{-1}\left\{ w_{t}(\bs_{t,i}^{*})-\mathbf{B}_{\bs_{t,i}^{*}}\bfw_{t,N_{t}(\bs_{t,i}^{*})}\right\} -\frac{b_{t}}{\sigma_{t}^{2}}\bigg],
\end{align*}
which is a $IG(a_{t}^{*},b_{t}^{*})$ distribution and 
\begin{align}
  a_{\sigma_{t}}^{*}& =a_{t}+\frac{1}{2}(n_{t}+n_{t}^{*}),\nonumber \\
 b_{\sigma_{t}}^{*} & =b_{t}+\frac{1}{2}\sum_{i=1}^{n_{t}}\left\{ w_{t}(\bs_{t,i})-\mathbf{B}_{\bs_{t,i}}\bfw_{t,N_{t}(\bs_{t,i})}\right\} ^{T}(\tilde{F}_{\bs_{t,i}})^{-1}\left\{ w_{t}(\bs_{t,i})-\mathbf{B}_{\bs_{t,i}}\bfw_{t,N_{t}(\bs_{t,i})}\right\}\nonumber \\
 & +\frac{1}{2}\sum_{i=1}^{n_{t}^{*}}\left\{ w_{t}(\bs_{t,i}^{*})-\mathbf{B}_{\bs_{t,i}^{*}}\bfw_{t,N_{t}(\bs_{t,i}^{*})}\right\} ^{T}(\tilde{F}_{\bs_{t,i}^{*}})^{-1}\left\{ w_{t}(\bs_{t,i}^{*})-\mathbf{B}_{\bs_{t,i}^{*}}\bfw_{t,N_{t}(\bs_{t,i}^{*})}\right\} . \label{eq.sigma_t}
\end{align}
For $\tau_{t}$, we have the full conditional distribution for each
level: 
\begin{align*}
p(\tau_{t}|\cdot)=IG(\tau_{t}|c_{t},d_{t})\times\prod_{i=1}^{n_{t}}N(\bfZ_{t}|\cdot).
\end{align*}
This structure gives us the inverse gamma distribution with: 
\begin{align}
  p(\tau_{t}|\cdot) & \sim IG(a_{\tau_{t}}^*,b_{\tau_{t}}^*)\nonumber \\
  a_{\tau_{t}}^{*} & =c_{t}+\frac{1}{2}n_{t},\nonumber \\
b_{\tau_{t}}^{*}  & =d_{t}+\frac{1}{2}\sum_{u\in\bfS_{t}}(z_{t}(\bs_{u})-\zeta_{t-1}(\bs_{u})y_{t-1}(\bs_{u})-\delta_{t}(\bs_{u}))^{2}. \label{eq.tau_t}
\end{align}

For $\bfgamma_{t}$, we have: 
\begin{align*}
p(\bfgamma_{t}|\cdot) & \propto N(\bfgamma_{t}|\boldsymbol{\mu}_{\bfgamma_{t}},\mathbf{V}_{\bfgamma_{t}})\times N(\bfZ_{t+1}|\mathbf{g}_{t}^{T}\bfgamma_{t}y_{t}(\bfS_{t+1})+\mathbf{h}_{t+1}\boldsymbol{\beta}_{t+1}+\bfw_{t+1},\tau_{t+1}\mathbf{I})\\
 & \propto\text{exp}\bigg\{-\frac{1}{2}(\bfgamma_{t}-\boldsymbol{\mu}_{\bfgamma_{t}})^{T}\mathbf{V}_{\bfgamma_{t}}^{-1}(\bfgamma_{t}-\boldsymbol{\mu}_{\bfgamma_{t}})\bigg\}\times\\
 & \quad\;\text{exp}\bigg[-\frac{1}{2\tau_{t+1}^{2}}\bigg\{ \bfZ_{t+1}-\mathbf{g}_{t}^{T}\bfgamma_{t}y_{t}(\bfS_{t+1})-\bfdelta_{t+1}\bigg\}^{T}\bigg\{ \bfZ_{t+1}-\mathbf{g}_{t}^{T}\bfgamma_{t}y_{t}(\bfS_{t+1})-\bfdelta_{t+1}\bigg\}\bigg]\\
 & \propto\text{exp}\bigg[-\frac{1}{2}\bfgamma_{t}^{T}\mathbf{V}_{\bfgamma_{t}}^{-1}\bfgamma_{t}- \frac{1}{2\tau_{t+1}^{2}}y_{t}(\bfS_{t+1})^{T}y_{t}(\bfS_{t+1})(\mathbf{g}_{t}^{T}\bfgamma_{t})^{T}(\mathbf{g}_{t}^{T}\bfgamma_{t})+ \nonumber \\
 &  \quad\;\left\{\boldsymbol{\mu}_{\bfgamma_{t}}^{T}\mathbf{V}_{\bfgamma_{t}}^{-1}+\frac{1}{\tau_{t+1}^{2}}\left[(\bfZ_{t+1}-\bfdelta_{t+1})^{T}y_{t}(\bfS_{t+1})\right]\mathbf{g}_{t}^{T}\right\}\bfgamma_{t}\bigg]
\end{align*}
so that: 
\begin{align}
 \bfgamma_{t}|\tilde{\bfy}_{t},\boldsymbol{\beta}_{t+1},\tau_{t+1},\bfZ_{t+1}  &  \sim N(\mathbf{V}_{\bfgamma_{t}}^{*}\boldsymbol{\mu}_{\bfgamma_{t}}^{*},\mathbf{V}_{\bfgamma_{t}}^{*}),\nonumber \\
  \boldsymbol{\mu}_{\bfgamma_{t}}^{*}  &  =\boldsymbol{\mu}_{\bfgamma_{t}}^{T}\mathbf{V}_{\bfgamma_{t}}^{-1}+\frac{1}{\tau_{t+1}^{2}}\left[(\bfZ_{t+1}-\bfdelta_{t+1})^{T}y_{t}(\bfS_{t+1})\right]\mathbf{g}_{t}^{T},\nonumber \\
 \mathbf{V}_{\bfgamma_{t}}^{*}  &  =\left(\mathbf{V}_{\bfgamma_{t}}^{-1}+\frac{1}{\tau_{t+1}^{2}}y_{t}(\bfS_{t+1})^{T}y_{t}(\bfS_{t+1}) \mathbf{g}_{t}\mathbf{g}_{t}^{T}\right)^{-1}. \label{eq.gamma_t}
\end{align}

For a new input location $s_{p}\not\in\boldsymbol{\tilde{S}}_{t}$, we have the predictive distribution of $w_{t}(s_{p})$:\allowdisplaybreaks
{\small{}{}{} 
\begin{align}
w_{t}(\bs_{p})\sim & N(V_{t,\bs_{p}}\mu_{t,\bs_{p}},V_{t,\bs_{p}}),\nonumber \\
V_{t,\bs_{p}}= & (\tau_{t}^{-2}+\tilde{F}_{t,\bs_{p}}^{-1})^{-1},\nonumber \\
\mu_{t,\bs_{p}}= & \tau_{t}^{-2}[z_{t}(\bs_{p})-\mathbf{h}_{t}^{T}(\bs_{p})\boldsymbol{\beta}_{t}-\zeta_{t-1}(\bs_{p})y_{t-1}(\bs_{p})]+\tilde{F}_{t,\bs_{p}}^{-1}\mathbf{\tilde{B}}_{t,\bs_{p}}\tilde{\bfw}_{t,\tilde{N}_{t}(\bs_{p})},\ \ t=1,2,\ldots,T-1, \label{eq.w_prediction}
\end{align}
}with $\mathbf{\tilde{B}}_{t,\bs_{p}}=\mathbf{C}_{\bs_{p},\tilde{N}_{t}(\bs_{p})}^T\mathbf{C}_{\tilde{N}_{t}(\bs_{p})}^{-1}$,
$\tilde{F}_{t,\bs_{p}}=\mathbf{C}(\bs_{p},\bs_{p})-\mathbf{C}_{\bs_{p},\tilde{N}_{t}(\bs_{p})}^T\mathbf{C}_{\tilde{N}_{t}(\bs_{p})}^{-1}\mathbf{C}_{\bs_{p},\tilde{N}_{t}(\bs_{p})}$,
$\tilde{N}_{t}(\bs_{p})$ is the m nearest neighbors in $\boldsymbol{\tilde{S}}_{t,<p}$,
and $\tilde{\bfw}_{t,\tilde{N}_{t}(\bs_{p})}$ is the corresponding nearest
neighbor subset of $\tilde{\bfw}_{t}$.

\section{Performance Metrics}\label{Metrics}

In the empirical comparisons, we used the following performance metrics: 
\begin{enumerate}
\item Root mean square prediction error (RMSPE) is defined as 
\[
\text{RMSPE}=\sqrt{\frac{1}{n}\sum_{i=1}^{n}(y_{i}^{\text{pred}}-y_{i}^{\text{obs}})^{2}}
\]
 where $y^{\text{obs}}$ is the observed value in test data-set and
$y_{i}^{\text{pred}}$ is the predicted value from the model. It measures
the accuracy of the prediction from model. Smaller values of RMSPE
indicate more a accurate model.
\item Nash-Sutcliffe model efficiency coefficient (NSME) is defined
as: 
\begin{gather*}
\text{NSME}=1-\frac{\sum_{i=1}^{n}(y_{i}^{\text{pred}}-y_{i}^{\text{obs}})^{2}}{\sum_{i=1}^{n}(y_{i}^{\text{obs}}-\Bar{y}^{\text{obs}})^{2}}
\end{gather*}
where $y^{\text{obs}}$ is the observed value in test data-set and
$y_{i}^{\text{pred}}$ is the predicted value from the model. NSME
gives the relative magnitude of the residual variance from data and
the model variance. NSME values closer to $1$ indicate that the model has
a better predictive performance. 
\item 95\% CVG is the coverage probability of 95\% equal tail prediction
interval. 95\% CVG values closer to $0.95$ indicate better prediction
performance for the model. 
\item 95\% ALCI is average length of 95\% equal tail prediction interval.
Smaller 95\% ALCI values indicate better prediction performance for
the model. 
\item Deviance Information Criterion (DIC) and the effective number of parameters of the model($p_D$) are defined as: 
\begin{align*}
  D(\theta) & =-2\text{log}(p(y|\theta))+C,\\
  p_{D} & =\overline{D(\theta)}-D(\Bar{\theta}),\\
 \text{DIC} & =p_{D}+\overline{D(\theta)}
\end{align*}
It is used in Bayesian model selection. Models with smaller DIC and
$p_{D}$ are preferable. 
\end{enumerate}

\end{document}